\documentclass[twocolumn,useAMS,usenatbib,usegraphicx]{mn2e}
\usepackage{fixltx2e}
\usepackage[utf8]{inputenc} 
\usepackage{amsmath}
\let\originalleft\left
\let\originalright\right
\renewcommand{\left}{\mathopen{}\mathclose\bgroup\originalleft}
\renewcommand{\right}{\aftergroup\egroup\originalright}
\def\zl{{z_\textsubscript{l}}}
\def\zi{{z_{\textsubscript{l}i}}}
\def\zs{{z_\textsubscript{s}}}
\def\dl{{D_\textsubscript{l}}}
\def\nf{{N\left(F\right)}}
\def\tmu{{\tau\left(\mu\right)}}
\def\pmu{{P\left(\mu\right)}}
\def\pth{{P\left(\theta\right)}}
\def\pzl{{P\left(\zl\right)}}
\def\fc{{F_\textsubscript{c}}}
\def\lcdm{{$\Lambda$CDM }}
\def\oml{{\Omega_\Lambda}}
\def\omm{{\Omega\textsubscript{m}}}
\def\dtzl{{d\tau/d\zl}}
\def\dtz2{{d\tau\left(\zl\right)/d\zl}}
\def\rs{{r_\textsubscript{s}}}
\def\rhos{{\rho_\textsubscript{s}}}
\def\nl{{N_\textsubscript{l}}}
\def\pmi{{P\left(\mu_i\right)}}
\def\pzi{{P\left(\zi\right)}}

\newcommand{\cardiff}{{School of Physics \& Astronomy, Cardiff University, 5 The Parade, Cardiff, CF24 3AA, Wales, United Kingdom}}
\usepackage{soul}
\title[Strong Gravitational Lens Statistics using the \emph{Herschel}-ATLAS]{Strong Gravitational Lens Statistics using the \emph{Herschel}-ATLAS}
\author[Short, Pearson, Coles \& Eales]{Jo Short, Elizabeth Pearson, Peter Coles, and Steve Eales\\ \cardiff \\}

\begin{document}

\maketitle

\begin{abstract}
The identification of strong gravitational lenses in large surveys has historically been a rather time consuming exercise. Early data from the \emph{Herschel} Astrophysical Terahertz Large Area Survey (\emph{Herschel}-ATLAS) demonstrate that lenses can be identified efficiently at submillimetre wavelengths using a simple flux criteria. Motivated by that development, this work considers the statistical properties of strong gravitational lens systems which have been, and will be, found by the \emph{Herschel}-ATLAS. Analytical models of lens statistics are tested with the current best estimates for the various model ingredients. These include the cosmological parameters, the mass function and the lens density profile, for which we consider the singular isothermal sphere (SIS) and the Navarro, Frenk \& White (NFW) approximations. The five lenses identified in the \emph{Herschel}-ATLAS Science Demonstration Phase suggest a SIS density profile is preferred, but cannot yet constrain $\oml$ to an accuracy comparable with other methods. The complete \emph{Herschel}-ATLAS data set should be sufficient to yield competitive constraints on $\oml$. Whilst this huge number of lenses has great potential for constraining cosmological parameters, they will be most powerful in constraining uncertainty in astrophysical processes. Further investigation is needed to fully exploit this unprecedented data set.

\end{abstract}

\section{Introduction}
Since the first strong gravitational lens was discovered in 1979 (\citeauthor{Walsh1979}), hundreds have been found and studied \citep{Treu2010}. As well as serendipitous discoveries, dedicated surveys have searched for lens candidates. The Cosmic Lens All-Sky Survey \citep{Browne2003} identified 22 new strong gravitational lenses from imaging in the radio waveband. The Sloan Lens ACS Survey \citep{Bolton2006,Auger2009} has so far confirmed 85 new gravitational lenses using spectroscopic data from the SDSS to identify candidates for follow-up with high-resolution HST imaging. The Cosmological Evolution Survey \citep{Faure2008,Jackson2008} identified 67 lenses from visual inspection of HST imaging. In addition there are projects, such as MUSCLES \citep{Jackson2011} and SQLS \citep{Inada2012}, that are looking for new lenses by reprocessing data from large surveys such as UKIDSS and SDSS. 

There has been much interest in studying gravitational lenses because they magnify sources that would otherwise be too faint or distant to see. Historically these lens systems have been difficult to identify in sufficiently large numbers to be statistically useful. Some lens searches involve painstaking `by eye' analysis of each object observed. Now new methodologies are being developed to improve the efficiency of finding lenses and to improve the conversion rate from candidate to confirmed lensed source. These include using colours \citep{Ofek2007}, spectroscopic redshifts \citep{Auger2009}, and other imaging processing techniques \citep{Marshall2008}.

For a rigorous statistical study of gravitational lenses, they should all come from a single survey. This is because the probability of identifying lensed sources varies between surveys due to the different instrumental sensitivities and selection effects. 

For sources observed in the submillimetre waveband it has long been thought that galaxy number counts should fall off rapidly with increasing flux density. This is because of the rapid evolution of proto-spheroidal galaxies into starburst galaxies \citep{Granato2004}. This drop in counts with flux using submillimetre observations was anticipated to yield an effective way of identifying lenses \citep{Blain1996,Negrello2007}. It has since been successfully demonstrated using the \emph{Herschel}-ATLAS data \citep{Negrello2010, Gonzalez-Nuevo2012}. Lensed sources are identifiable much more easily because the lensing magnification pushes them into higher flux densities which have intrinsically low source populations.

The \emph{Herschel} Space Observatory, launched in 2009, is the only space-based observatory to cover a spectral range from the far infrared to the submillimetre. It provides a unique window through which to study large scale structure. The \emph{Herschel} Astrophysical Terahertz Large Area Survey \citep[\emph{Herschel}-ATLAS]{Eales2010} will cover 550 deg$^2$ of the sky making it the \emph{Herschel} survey with the largest area. \cite{Negrello2010} showed that strong gravitational lenses could be identified with almost 100\% efficiency by selecting sources which have 500$\mu$m flux densities greater than 100mJy. \cite{Gonzalez-Nuevo2012} further improved on this technique by refining the flux selection to incorporate other \emph{Herschel} wavebands and optical/infrared data from the VIKING \citep{Fleuren2012,Sutherland2012} survey to look for counterparts. This technique is forecast to yield ten times more lenses than that by \citeauthor{Negrello2010}. The expected efficiency of 70\% is lower that that achieved by \citeauthor{Negrello2010}, but still impressive. Using these techniques, the \emph{Herschel}-ATLAS team expect to find as many as a thousand strongly lensed candidates.

We emphasise that this article considers the likelihood of \emph{strong} gravitational lenses which are defined as lensed systems with multiply imaged sources. Recent articles on the effect of lensing on the source number counts \citep{Lima2010, Wardlow2012} do not consider strong lensing in the sense that sources are multiply imaged, just that they are highly magnified. Note, this results in the lens probability distribution assuming a NFW density profile becoming much more similar to that of the SIS density profile than is seen in our analysis. Looking to the future, `strongly' lensed sources are being selected by their excess submillimetre flux, or effectively by their magnification, with limited optical follow-up to confirm multiple imaging. Since requiring optical follow-up may limit the number of confirmed sources, it may in future turn out to be statistically favourable to relax the multiple image requirement to gain a statistical advantage.  However our approach, looking at strong gravitational lens systems with multiple images, is used to be consistent with the nature of the sources identified by \cite{Negrello2010}.

The statistical properties of gravitationally lensed sources were first considered in the 1980s \citep{Turner1984} and later applied as a possible way of constraining the cosmological model \citep{Li2002,Mitchell2005,Zhang2009,Dobke2009}. With the prospect of a thousand lenses being identified by \emph{Herschel}-ATLAS, we take a fresh look at the analytical theory behind predicting strong gravitational lens statistics in Sections \ref{sec1a} and \ref{sec1b}. In Section \ref{sec2} the lenses identified by \cite{Negrello2010} are compared against the analytical predictions, and then in Section \ref{sec3} we consider the parameter constraints possible with the full \emph{Herschel}-ATLAS data set.

Unless otherwise stated we use cosmological parameters: $\omm$ = 0.27, $\oml$ = 0.73, $\Omega_\textsubscript{b}$ = 0.046, h = 0.7, n = 0.97, and $\sigma_8$ = 0.81 \citep{Komatsu2011}.

\section{Lens Statistics: the theory}
\label{sec1a}
A source is defined to be strongly gravitationally lensed when it has been multiply imaged due to the gravitational effects of mass between the source and the observer. The probability of a source being lensed is known as the optical depth ($\tau$). It depends on three main components: the cosmology which determines the co-moving volume element at a given redshift; the normalised mass function ($n$) which describes the number density of halos; and the lensing cross-section ($\sigma$), which is the area in the lens plane where the separation between the lens and source is sufficiently small for strong lensing to occur:
\begin{eqnarray}
 \frac{d\tau\left(\zs, \zl\right)}{d\zl} = \int\limits^\infty_0 \, \frac{d\dl}{d\zl} \, \left(1+\zl\right)^3 \, n\left(m,\zl\right) \, \sigma\left(m,\zl\right)\, dm ,
 \label{eqn1}
\end{eqnarray}
where $\dl$ is the angular diameter distance to the lens, $\zl$ is the lens redshift, $\zs$ is the source redshift, and $m$ is the virial mass of the lens \citep{Schneider1993}. This is explained in detail in the Appendix, where we describe how to calculate each of the different components. As well as the dependence on the lensing redshift $\zl$, the dependence on other parameters such as the total magnification of the lensed source or the angular separation of the images can be calculated.

When comparing the theoretical predication for the distribution of strong lenses with observational surveys, another factor that needs to be taken into account is magnification bias \citep{Turner1984}. 

\subsection{Magnification Bias}
\label{ApMagBias}
Magnification bias is a correction to account for that fact that lensed sources are magnified and are consequently observed to be brighter than the population from which they are drawn. Therefore sources which are lensed come from a population with a lower flux. Since there are generally more fainter sources than there are brighter sources, magnification bias almost always increases the chance of finding a lens compared to if the lenses came from the same flux population at which they are observed. 

The magnification bias at a particular flux is calculated as the ratio of the number of lenses which will be magnified to that flux over the number of objects which intrinsically have that flux times the probability of lensing:
\begin{eqnarray}
 B\left(F\right) = \frac{ \int\limits^\infty_1 \, \tmu \, N\left(\frac{F}{\mu}\right) \, d\mu }{ \tau \, \nf} 
                 = \frac{ \int\limits^\infty_1 \, \pmu \, N\left(\frac{F}{\mu}\right) \, d\mu }{ \nf },
\label{biasf}
\end{eqnarray}
where $\pmu = \tmu/\tau$ is the probability of a magnification $\mu$ given that the source has been lensed\footnote{Note that we do not here include a $\mu^{-1}$ factor as in \cite{Turner1984}. As detailed in \cite{Fukugita1991} since both $\tau$ and $\tmu$ are considered in the same plane this `correction' factor is not required}.

The overall bias is then calculated as the average of Eq. \ref{biasf} over all fluxes:
\begin{eqnarray}
 B = \frac{{\int\limits^\infty_F}_\textsubscript{lim} \int\limits^\infty_1 \, \pmu \, N\left(\frac{F}{\mu}\right) \, d\mu \, dF}{{\int\limits^\infty_F}_\textsubscript{lim} \, \nf \, dF},
\label{bias1}
\end{eqnarray}
where $F_\textsubscript{lim}$ is the flux limit of the observational survey.

Generally the flux distribution is assumed to take the form of a simple power law $\nf = N_0 \, \left(F/F_0\right)^{-\beta}$ and so Eq. \ref{bias1} then reduces to $B =  \int \, \tmu) \, \mu^{\beta} \, d\mu$. However, for this particular survey the slope $\beta$ is so steep that $\mu^{\beta}$ grows faster than $\tmu$ decreases and so this integral does not converge. But assuming that this power law approximation for $\nf$ holds for infinitely small fluxes is clearly incorrect; there will come a point where the flux counts fall off. To account for this we introduce an exponential term that stops the growth of $\nf$ below a given flux $\fc$ meaning that $\nf$ will return to zero as $F$ tends to zero. We use $\nf = N_0 \left(F/F_0\right)^{-\beta} \exp\left(-\left(F/\fc\right)^2\right)$ and the magnification bias becomes:
\begin{eqnarray}
B = \frac{{ \int\limits^\infty_1 \, \pmu \, \mu^{\beta} \, \int\limits^\infty_F}_\textsubscript{lim} \, F^{-\beta} \, \exp\left(-\left(F/ \mu \fc\right)^2\right)  \, dF\, d\mu}{{\int\limits^\infty_F}_\textsubscript{lim} \, F^{-\beta} \, \exp\left(-\left(F/\fc\right)^2\right)\, dF}.
\label{bias2}
\end{eqnarray}
In this work we simplify the above expression by setting $F_\textsubscript{lim}$ to 0. Eq. \ref{bias2} then reduces to:
\begin{eqnarray}
B = \int\limits^\infty_1 \, \pmu \, \mu \, d\mu,
\end{eqnarray}
which intuitively makes sense as it is just the average magnification. This work assumes that $F_\textsubscript{lim}$ = 0.
\\

Whilst Eq. \ref{eqn1} is relatively simple, it is more complex to implement in practice due to the uncertainties associated with each of the constituent components. In the rest of this section we explore the current best estimates for each of the components and consider the likely impact of uncertainty in each of them on the probability distributions for observing gravitational lenses.

\section{Lens Statistics: Demonstrating the theory}
\label{sec1b}

In this section the different components of the lensing optical depth are considered. We discuss the current best estimates for each component and demonstrate the corresponding effects of uncertainty on the optical depth.

\subsection{Dark Energy Density}
The community is, in general, confident in the \lcdm cosmological model refined from CMB \citep{Bennett2003}, BAO \citep{Percival2010} and supernovae \citep{Riess1998} observations. However it is still worth considering how the different cosmological parameters would affect these lensing statistics. If variations in the underlying cosmology significantly affect the lensing optical depth then the statistics of gravitational lenses could provide another complementary test of the cosmology. On the other hand, if the cosmological parameters have very little impact on the lensing statistics then we can be confident that any assumptions we draw are not dependent on the accuracy of the \lcdm assumption.

\begin{figure}
 \begin{center}
   \includegraphics[width=\linewidth]{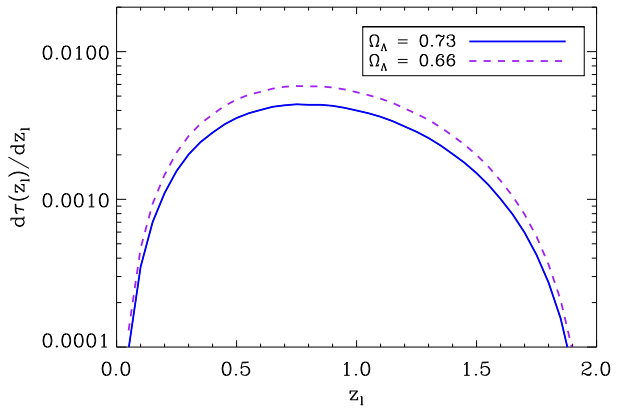}
  \caption{The differential lensing optical depth as a function of lens redshift for $\oml = 0.66$ (purple dashed line) and 0.7 (blue solid line). This assumes a source redshift $\zs=2$, SIS lens density profile and a \citet{Tinker2008}  mass function. The increase in the comoving volume element and decrease the density perturbation growth rate combine so that the overall optical depth decreases with increasing $\oml$. } 
 \label{figTauCosP}
 \end{center}
\end{figure}

The dark energy density ($\oml$) contributes to the lensing optical depth (described in Eq. \ref{eqn1} and the Appendix) by increasing the comoving volume the light from a source travels through. This makes it more likely that a source will encounter an object and be lensed. If we assume a flat universe, this increase in $\oml$ is balanced by a reduction in the matter density ($\omm$). This decrease in $\omm$ will reduce the density perturbation growth rate and hence decrease the mean mass in the mass function. Figure \ref{figTauCosP} shows an example of the differential lensing optical depth as a function of lens redshift for two different dark energy densities ($\oml = 0.66$ and 0.73). The increase in the comoving volume element and decrease in the density perturbation growth rate combine so that the overall optical depth decreases with increasing $\oml$. In this example a 10\% decrease in $\oml$ results in a factor of two increase in the differential lensing optical depth. There is also a slight bias to lower $\zl$ with increasing $\oml$. The shape of the differential lensing optical depth does not significantly change as a function of $\theta$ and $\mu$ (not shown).

Previous work has used results from the CLASS \citep{Browne2003} survey to constrain $\oml$. The CLASS data set, which includes 13 lenses, was used to constrain the total dark energy density $\oml$ for a flat Universe \citep{Chae2002, Mitchell2005} and the dark energy equation of state parameter \citep{Zhang2009}. For illustrative purposes, here we consider just varying the dark energy density for a flat cosmology. There are many other cosmological parameters which could be constrained such as the curvature. This would affect the apparent angular separation of the lens images. A positive curvature would make the angular separations appear smaller than they would in a flat universe \citep{Park1997}. With the large number of lenses anticipated in the future it may also be possible to constrain modified gravity \citep{Zhao2011}.

\subsection{Lens Distributions}
Understanding the lens distribution is important as lensing is a gravitational effect and can tell us about the distribution of all matter not just baryonic matter. The distribution of matter over-densities, or halos, is frequently approximated using an analytical approximation called the mass function.

The mass function $n\left(m,z\right)$ (sometimes written as $dn/\,dm$) predicts the differential number of halos of a given mass as a function of redshift. It is frequently approximated using the approach by \cite{Press1974}:
\begin{eqnarray}
  \frac{m^2 \, n\left(m,\zl\right)}{\bar\rho} \, \frac{dm}{m} = \nu \, f\left(\nu\right)\, \frac{d\nu}{\nu},
\end{eqnarray}
where $\bar\rho$ is the average matter density,  $\nu\left(m,\zl\right) = \delta_\textsubscript{c}^2/\sigma^2$, $\delta_\textsubscript{c}$ is the critical density required for spherical collapse and $\sigma^2$ is the variance in the linear density fluctuation field (see also Appendix \ref{secApMassFunc}). There have been a number of estimates proposed for $f\left(\nu\right)$. \citet{Press1974} derived an approximation based on theoretical arguments. This was later shown to under/over predict the abundance of halos of high/low mass relative to N-body simulations \citep{Sheth1999}. Improved estimates have been proposed \citep{Jenkins2001, Warren2006}, most recently by \cite{Tinker2008} whose fit introduced a specific dependence on redshift.

\begin{figure}
 \begin{center}
   \includegraphics[width=\linewidth]{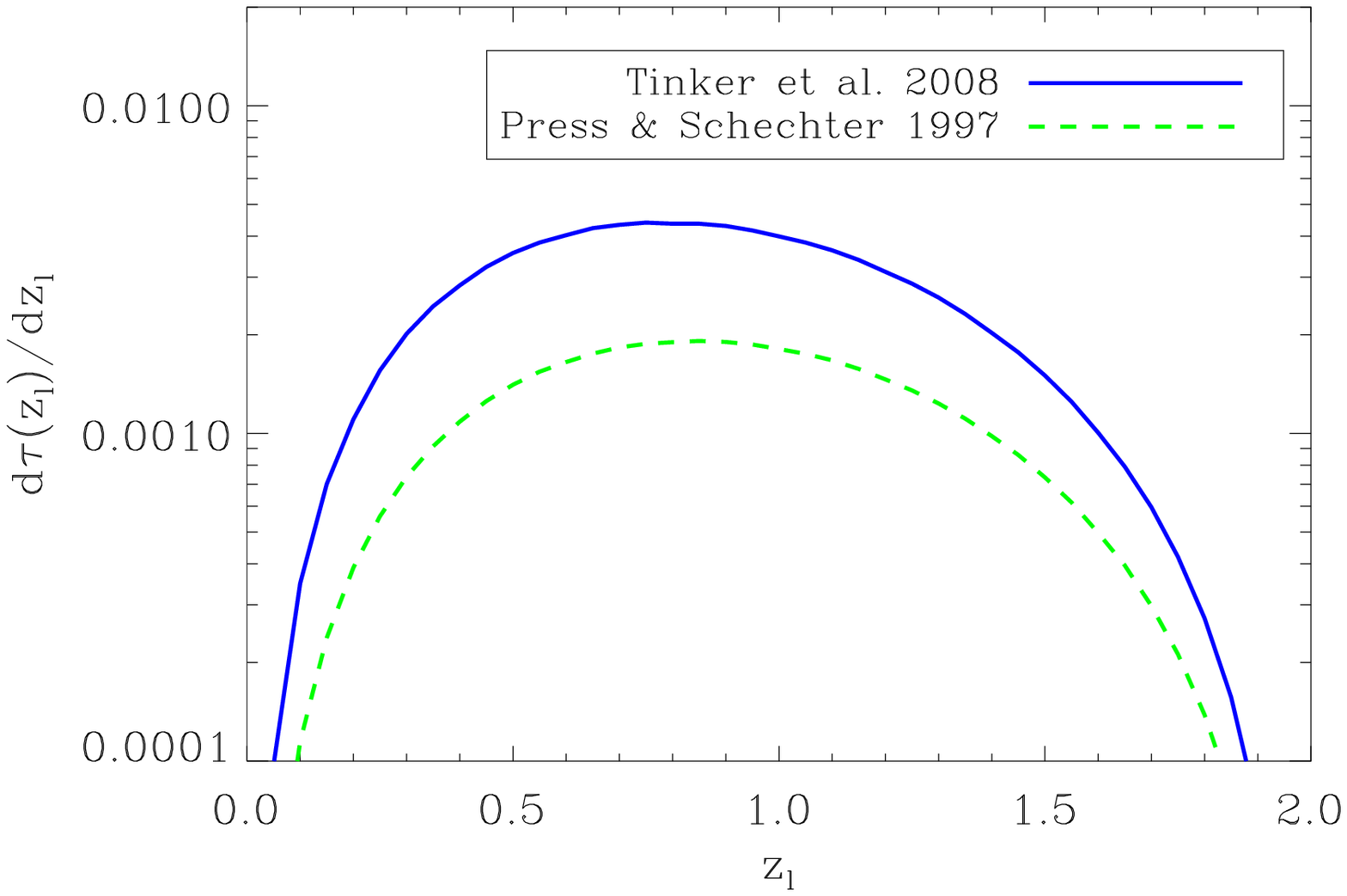}
   \includegraphics[width=\linewidth]{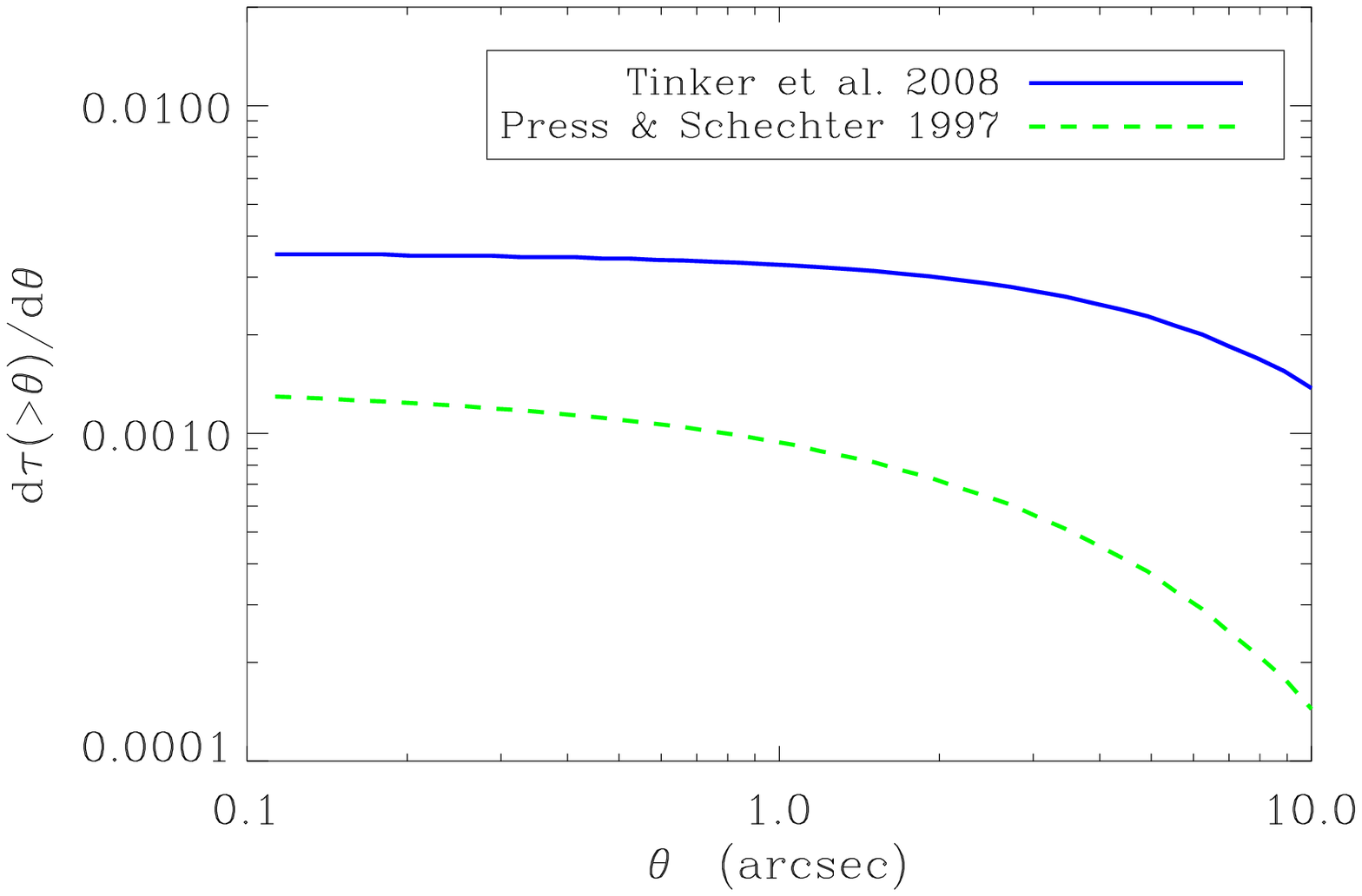}
  \caption{The differential lensing optical depth as a function of redshift for $\zs=2$ and a SIS lens profile. The green dashed line is for the \citeauthor{Press1974} mass function, which is compared to the \citeauthor{Tinker2008} mass function (blue solid line). The \citeauthor{Press1974} mass function underestimates the lensing optical depth compared to the \citeauthor{Tinker2008} mass function.  }
 \label{figTauMF}
 \end{center}
\end{figure}

So how does uncertainty in the mass function $n\left(m,\zl\right)$ affect the differential lensing optical depth? Figure \ref{figTauMF} shows the differential lensing optical depth for both the \citet{Press1974} and \citet{Tinker2008}  approximations for the mass function. The \citeauthor{Press1974} mass function estimates lower values of $\dtzl$ than that the \citeauthor{Tinker2008} mass function.  The  shape of $\dtzl$ tends to favour: larger $\zl$ for the \citeauthor{Press1974} mass function compared to the \citeauthor{Tinker2008} mass function; and smaller $\theta$ for the \citeauthor{Press1974} mass function compared to the \citeauthor{Tinker2008} mass function. This is because the \citeauthor{Press1974} mass function predicts a larger $n$ at small $m$, and a smaller $n$ at high $m$, compared to the \citeauthor{Tinker2008} mass function. Integrating over all $m$ results in a smaller $\dtzl$ and $d\tau/d\theta$ for the \citeauthor{Press1974} mass function because the higher mass halos contribute most to the lensing optical depth. $d\tmu/d\mu$ maintains the same shape for both mass functions (not shown).
 
Whilst it is theoretically possible to generate a lensed source with a lens made purely of dark matter, the way in which candidate lens sources are confirmed relies on follow-up imaging to identify a lens candidate. Without an observed lens candidate it is likely the source would be rejected as a potential lens. This is not something that has been accounted for in this work. The lensing optical depth is approximated with a mass function for the total matter distribution. This is worth noting as a potential source of error.

\subsection{Halo Density Profile}
\label{ch1_profile}
Eq. \ref{eqn1} is most sensitive to the lens density profile component. The simplest model for a halo density profile that can explain the flat rotation curves of galaxies is a singular isothermal sphere (SIS). This has a density profile given by:
\begin{eqnarray}   \rho\left(r\right) = \frac{v^2}{2\pi \, G \, r^2}, \end{eqnarray}
where $v$ is the velocity dispersion  (see Appendix \ref{secSIS}).  This profile has been successful in reproducing a number of features of gravitational lensing \citep{Auger2009}. However the profiles of dark matter halos generated via numerical simulations are better fit by the profile proposed by \citet[NFW]{Navarro1997}:
\begin{eqnarray}
\rho\left(r\right)=\frac{\rhos}{\left(\frac{r}{\rs}\right)\left(1 + \frac{r}{\rs}\right)^2},
\end{eqnarray}
where $\rs$ is the scale radius and $\rhos$ is the density at $\rs$ (see Appendix \ref{secNFW}). Neither the SIS or NFW density profiles is probably realistic for a lens made of a dark matter halo and baryon component. Here we use them for illustrative purposes to demonstrate the scale of uncertainty and its significant effect in the resulting optical depth.

SIS profiles produce more strong gravitational lenses than the NFW profile. This is because they have a steeper inner density gradient ($\rho \propto r^{-2}$ as opposed to $\rho \propto r^{-1}$ for the NFW) and steeper inner profiles are more effective at producing multiply imaged lenses \citep{Li2002}. NFW profiles produce lenses with higher mean magnifications than those for the SIS . This is because lenses that are less concentrated (such as the NFW) require the source and lens to be very closely aligned for strong lensing to result. It is this close alignment that results in a higher magnification \citep{Wyithe2001}. In other words lenses that are more concentrated (such as the SIS) do not need the lens and source to be so closely aligned for multiple imaging to occur and this results in many more lower magnifications. 

\begin{figure}
 \begin{center}
   \includegraphics[width=\linewidth]{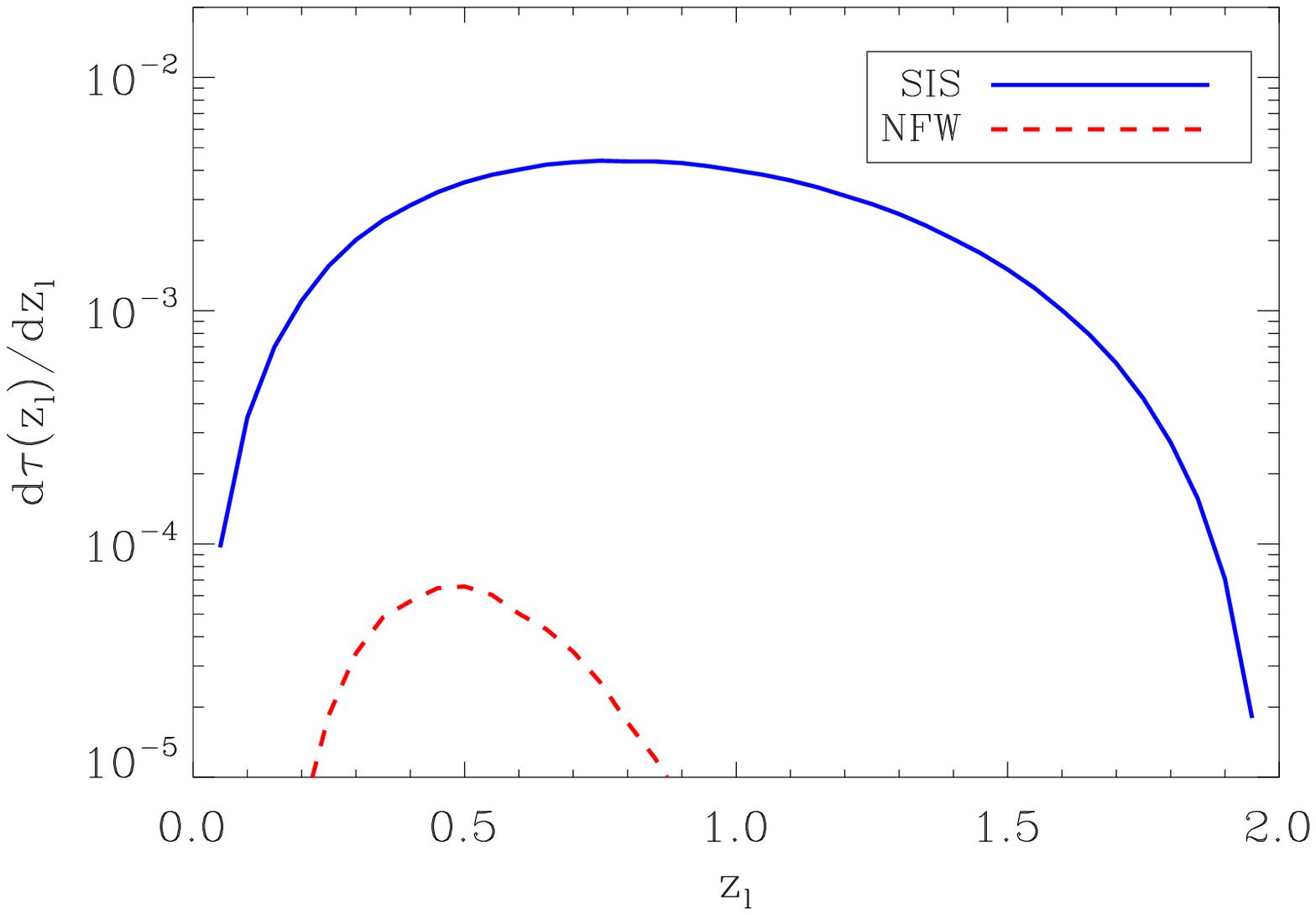}
   \includegraphics[width=\linewidth]{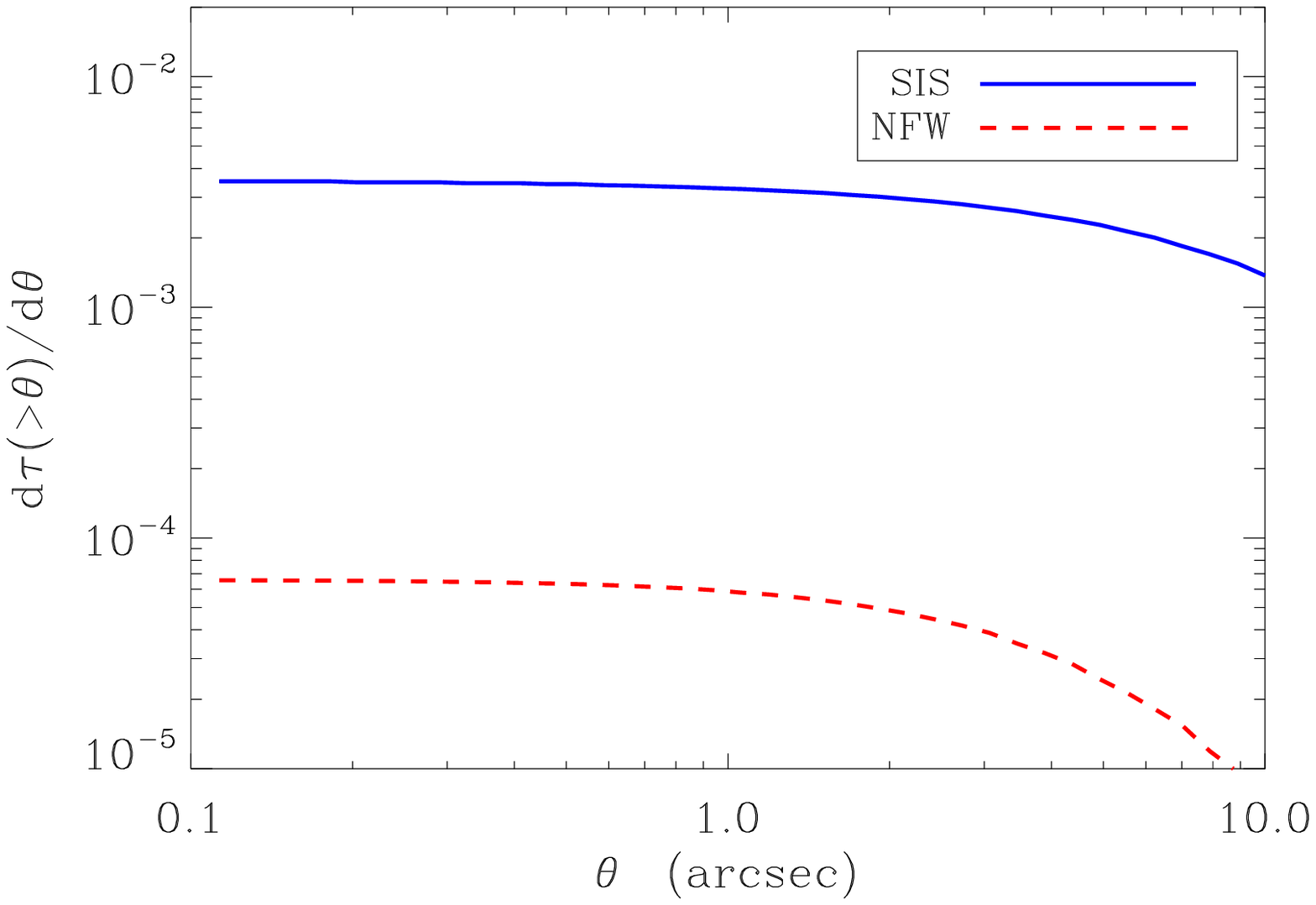}
   \includegraphics[width=\linewidth]{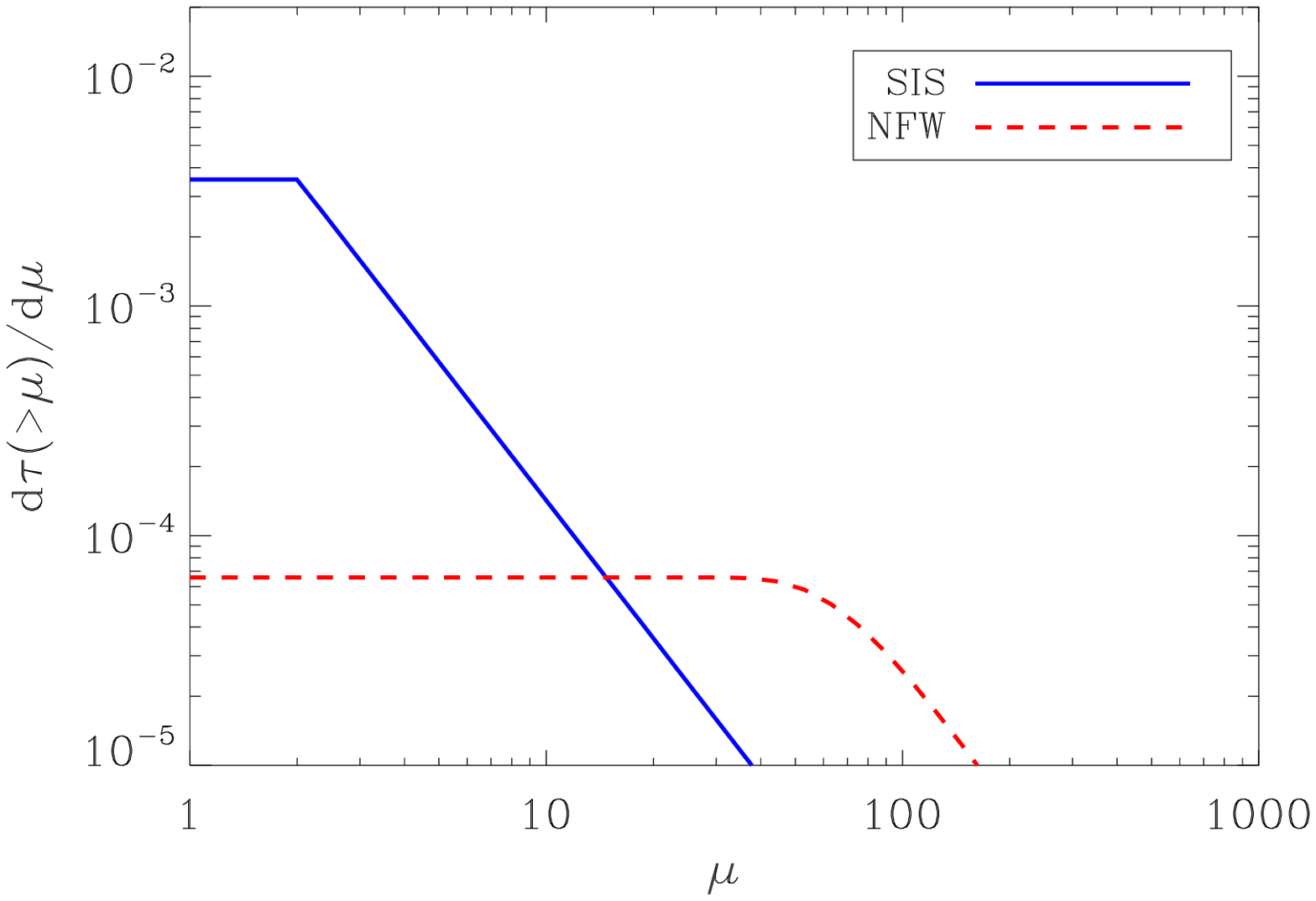}
    \caption{The differential lensing optical depth for $\zs=2.0$ and a \citeauthor{Tinker2008} mass function. When not varying the lensing redshift is fixed to $\zl = 0.5$. The solid blue line shows the SIS compared to the NFW (red dashed line) lens profiles. The SIS profile produces more lenses than the NFW profile because it has a steeper inner density gradient.}
 \label{figTaun10}
 \end{center}
\end{figure}

Figure \ref{figTaun10} compares the differential optical depth for the SIS and NFW halo profiles. As well as demonstrating the known results for $d\tau/d\mu$ discussed above, other effects are seen for $\dtzl$ and $d\tau/d\theta$. The NFW profile tends to predict more small-separation lenses and fewer large-separation lenses than the SIS profile. Also the mean lens redshift is much smaller for the NFW density profile than the SIS.

These results confirm that the theoretical strong gravitational lensing probabilities are highly sensitive to the density gradient near the centre of dark matter halos \citep{Li2002}. This is an area of tension in current observations \citep{Comerford2006, Shu2008}. High resolution observations of galaxy rotation curves suggest halo density profiles with a flat central core. However, studies of  gravitational lensing statistics suggest a steep, or `cuspy', density profile for halo cores \citep{Chen2008}.

A potential resolution for this is that the profiles of the gravitational lens population are biased with respect to the general population \citep{Mandelbaum2009}; alternative profiles have also been suggested. Following \cite{Navarro1997}, higher resolution numerical simulations \citep{Moore1999} suggested that the \textit{central} density gradient should lie somewhere in between that of the NFW ($\rho \propto r^{-1}$) and SIS ($\rho \propto r^{-2}$) profiles. 
It may be that parameterising the slope of the density profile with an Einasto \citep{Retana-Montenegro2011} or a S\'{e}rsic \citep{Mandelbaum2009,Eliasdottir2007} profile would be more appropriate.
 
Alternatively, CLASS found a number of lenses at low separation, but none at higher $\theta$, which suggested a combination of NFW and SIS density profiles. They proposed that if the mass of the lens is greater than a given cut-off $m_\textsubscript{c}$ then the lens density profile is approximated using a NFW approximation, else a SIS profile is used \citep{Li2002}. This was not motivated by direct observations/simulations of halo profiles, but it has been shown to provide a reasonable fit to lensing statistics for a mass cut-off scale of $m_\textsubscript{c} = 10^{13} M_\odot h^{-1}$ \citep{Zhang2009}.

The NFW halo density profile could well be accurate but the halo concentration could be incorrect. The \textit{concentration} of the halo (defined as $c = r_\textsuperscript{vir} / \rs$) has been studied extensively over the last decade. On average, the median halo concentration decreases with increasing mass and redshift; this can be approximated using analytical expressions \citep{Navarro1997,Bullock2001, Eke2001}.

All these suggestions assume a spherically symmetric lens, but realistically the halos are more likely to be irregularly shaped with a spherically symmetric average. Some way of accounting for this variation may well be required to provide a consistent explanation across dark matter simulations and lensing studies.

\subsection{Source Distributions}
The analytical calculation of the lensing optical depth in Eq. \ref{eqn1} only describes the probability of a source being lensed. It does not account for the probability of a source being at a particular redshift in the first place. So if you wanted to estimate the total optical depth this source redshift distribution needs to be measured observationally. 

For the \emph{Herschel} sources it is generally not possible to measure accurate spectroscopic redshifts due to the lack of emission lines in that band and the low angular resolution. Much effort has been put into deriving accurate source redshift distributions for the \emph{Herschel} source catalogues via other methods such as matching sources to optical counterparts with spectroscopic redshifts \citep{Smith2011}, CO line spectroscopy follow-up \citep{Harris2012}, and using models based on average galaxy SEDs \citep{Amblard2010, Lapi2011, Pearson2012}. All these methods have their own biases/constraints, so a potential test that could be done with this lens work would be to fix all the other components and investigate the source distribution.
\\
\\

In this section we have seen that the lensing statistics are sensitive to both astrophysical and cosmological parameters. The lensing optical depth is most sensitive to the lens density profile, both in magnitude and as a function of all $\zl$, $\mu$ and $\theta$. The mass function has little effect on the optical depth as a function of $\mu$, but does for $\zl$ and $\theta$. Of all the parameters considered, the optical depth was least sensitive to the dark energy density. It should be noted that before competitive constraints on the cosmological parameters can be achieved, the uncertainty in the properties of the lensing galaxies needs to be constrained.

\section{Lens Statistics: \emph{Herschel}-ATLAS SDP}
\label{sec2}
Five new lensed sources were discovered using the flux criteria technique on the \emph{Herschel}-ATLAS Science Demonstration Phase (SDP) data \citep{Negrello2010}. Table \ref{tab1} summarises the estimated source and lens redshifts for those lenses. 
\begin{table}
  \begin{center}
  \vspace{+0.5cm}
  \begin{tabular}{c|ccccc}
    \hline 
    ID & $\zl$ & $\zs$  & $\mu$ \\
    \hline
    \hline
    9  & 0.679$^{\pm0.057}$ &   1.577$^{\pm0.008}$  &$-$  \\
    11 & 0.7932$^{\pm0.0012}$ & 1.786$^{\pm0.005}$  & $-$\\
    17 & 0.9435$^{\pm0.0009}$  & 2.308$^{\pm0.011}$ & $-$ \\
    81 & 0.2999$^{\pm0.0002}$ & 3.042$^{\pm0.001}$&  25$\pm$7 \\
    130& 0.2201$^{\pm0.002}$& 2.6260$^{\pm0.0003}$ & 6$\pm$1 \\
    \hline
    \end{tabular}
    \caption{Estimates of the source ($\zs$) and lens ($\zl$) redshifts and magnification ($\mu$) for the five lenses discovered in the \emph{Herschel}-ATLAS SDP \citep{Negrello2010}. The quoted errors correspond to a 68$\%$ confidence interval.}  \label{tab1}
  \end{center}
\end{table}
Whilst this is probably not enough lenses with which to do convincing statistics, we can still consider whether or not these observations are consistent with the analytical models discussed in Section \ref{sec1a}.

Here we consider both the absolute differential optical depth as a function of lensing redshift $\dtz2$ and the conditional differential optical depth\footnote{Note here we consider $d\tmu/d\mu = [d\tau(>\mu+d\mu)-d\tau(>\mu)]/d\mu$ not $d\tau(>\mu)/d\mu$ as in Section \ref{sec1b}.}: 
\begin{eqnarray*}
\pmu = \frac{1}{\tau} \frac{d\tmu}{d\mu} \textnormal{ and } \pzl = \frac{1}{\tau} \frac{d\tau(\zl)}{d\zl} ,
\end{eqnarray*}
which describe the probability of a lens being at a certain $\mu$ or $\zl$ given that lensing has occurred. We do not consider $\pth$ here because estimates are not yet available for the image separations of the \emph{Herschel}-ATLAS SDP lenses. 

By choosing to consider the conditional probability, rather than the absolute probability, the calculation is simplified.  The magnification bias (see Section \ref{ApMagBias}) does not need to be calculated as the conditional probability is effectively the `normalised' optical depth for fixed $\zs$. The conditional probability is therefore the expected distribution of the lenses as a function of $\mu$ or $\zl$ given that lensing has occurred.

We start by looking to constrain the lens density profile since this is the component to which the statistics of strong lenses appeared to be most sensitive in Section \ref{ch1_profile}. We start by considering the conditional probability as a function of magnification $\pmu$ and redshift $\pzl$ to constrain the lens density profile between the NFW and SIS.

\subsection{Constraining the Lens Density Profile with $\pmu$} 
\label{sec2a}
\begin{figure}
 \begin{center}
  \includegraphics[width=\linewidth]{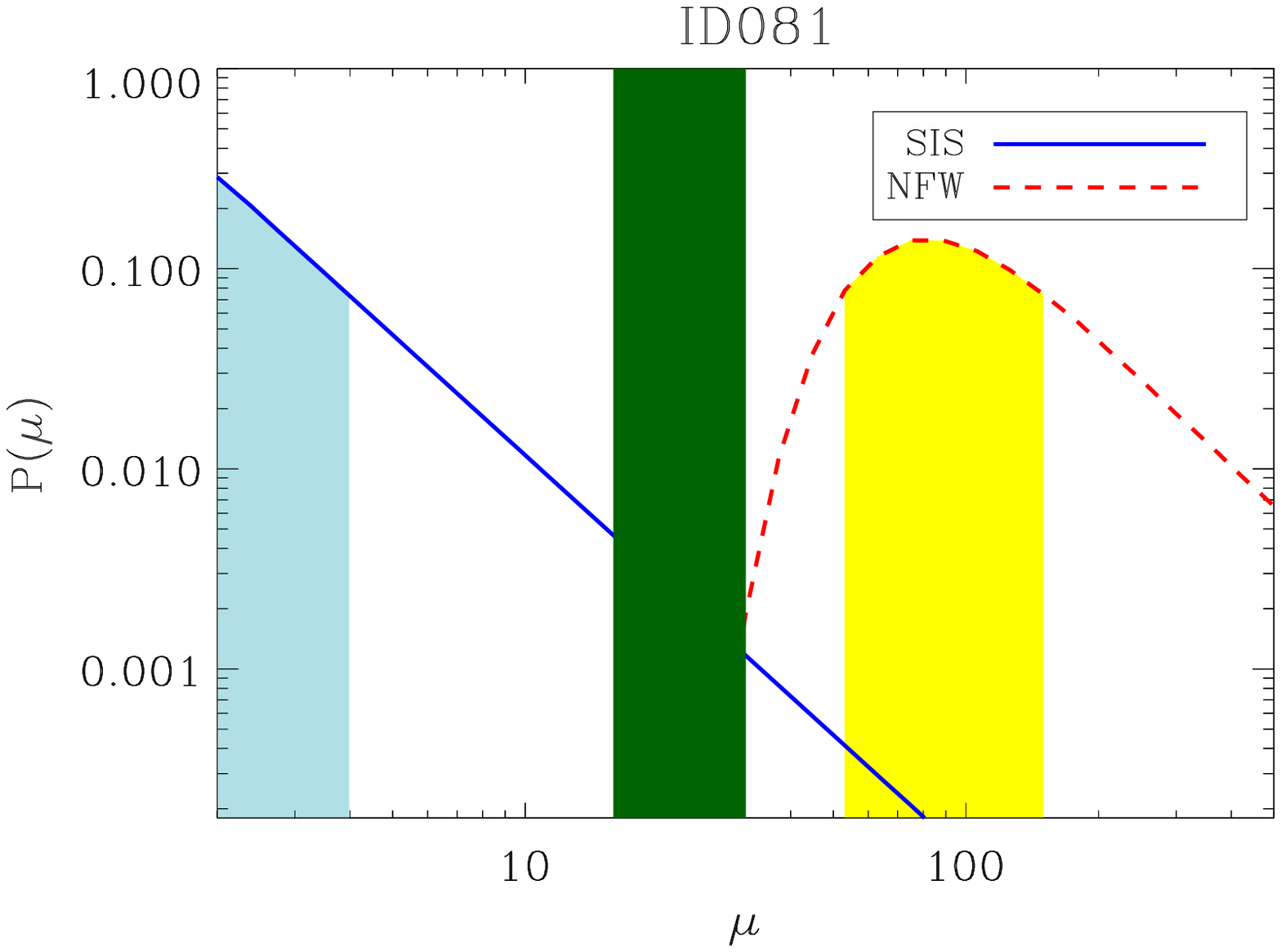}
  \includegraphics[width=\linewidth]{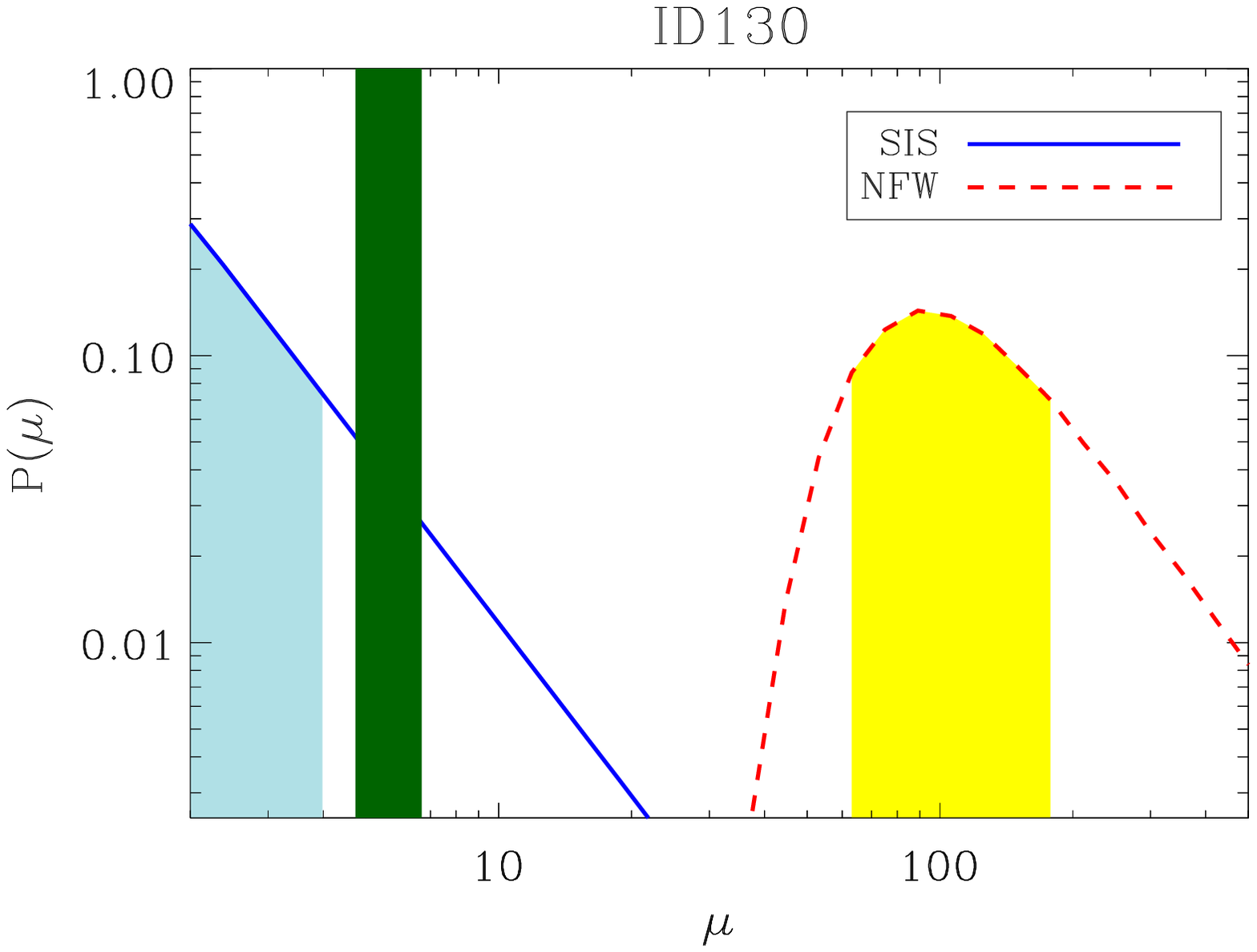}
  \caption{Conditional probability of finding a lens at a certain magnification given lensing has occurred, for source and lens redshifts in Table \ref{tab1}. Solid blue lines use the SIS lens density profile; dashed red lines use the NFW lens density profile. Shading indicates 68\% confidence intervals. Solid green vertical lines indicate the estimated magnification of gravitational lenses ID081 and ID130 found by \citet{Negrello2010}. }
 \label{figTauMag}
 \end{center}
\end{figure}

Magnifications are available for the \emph{Herschel}-ATLAS SDP lenses ID081 and ID130 \citep{Negrello2010}. Theoretical distributions of the conditional probability as a function of magnification were calculated as described in the Appendix. They were calculated assuming a \citeauthor{Tinker2008} mass function, and a standard \lcdm cosmology. The source and lens redshifts are as per the estimates in Table \ref{tab1}. The resulting $\pmu$ are shown in Figure \ref{figTauMag} for the two different estimates of the lens density profile (SIS and NFW), along with the two observational estimates of the magnifications for ID081 and ID130.
This shows that the magnification of ID081 source is within the 99\% confidence internal of the theoretical probability found for both the SIS and NFW density profiles. The magnification for the ID130 source is consistent with the model with the SIS density profile at a 95\% level, but inconsistent with the model with the NFW profile at over a 99\% level. These results appear to favour the SIS density profile. 

One way of quantifying the combination of the ID081 and ID130 results to choose between the NFW and SIS options is to use a likelihood. The likelihood of the data given the model parameters is defined as:
\begin{eqnarray}
 \mathcal{L}\left(\mu_i \, \Big | \, \rho_\textsubscript{SIS/NFW}\right) = \prod\limits^{\nl}_{i=1} \, \pmi ,
\label{eqn2}
\end{eqnarray}
where $\nl=2$ is the number of lensed sources being considered. To give the most favourable result for each of the  lens models, the highest probability associated with the range of observed magnification is chosen for each $\pmi$. Calculating this gives  $\mathcal{L} \left( \mu_i \, | \, \rho_\textsubscript{SIS}\right) = 3 \times 10^{-3}$ and  $\mathcal{L} \left( \mu_i \, | \, \rho_\textsubscript{NFW}\right) < 5 \times 10^{-6}$ which confirms that a SIS density profile is preferred overall. 

The observational estimates for the magnification in Figure \ref{figTauMag} are consistently higher than those preferred by the SIS lens density profile. We note that magnifications close to 1 will be much harder to identify observationally than higher magnifications. Therefore we would expect that the observational values would be biased against magnifications close to $\mu=1$. This suggests that the SIS likelihood is actually higher. Overall we see the \emph{Herschel}-ATLAS SDP lens magnifications show that the SIS density profile is much preferred over the NFW density profile.

\subsection{Constraining the Lens Density Profile with $\pzl$}
\label{sec2b}
In Section \ref{sec1b} we saw that the lens density profile strongly impacts $\pzl$ and $\pth$ as well as $\pmu$. Data is not yet available for the angular separation of the lensed images, therefore we just look at $\pzl$ here. There were only two data points available for the lens magnifications but lens redshift estimates are available for all of the \emph{Herschel}-ATLAS SDP lenses.
\begin{figure*}
 \begin{center}
  \includegraphics[width=0.49\linewidth]{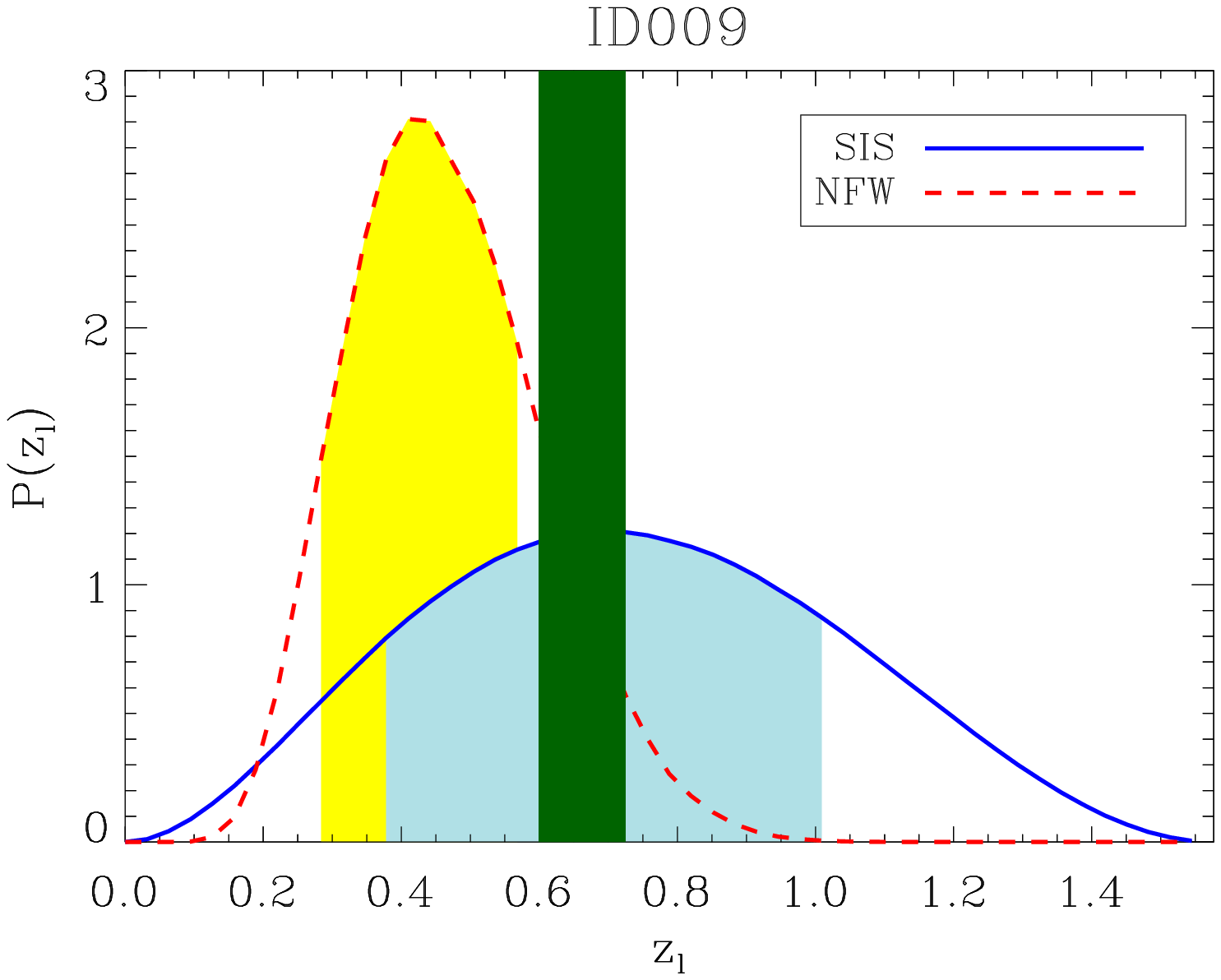}
  \includegraphics[width=0.49\linewidth]{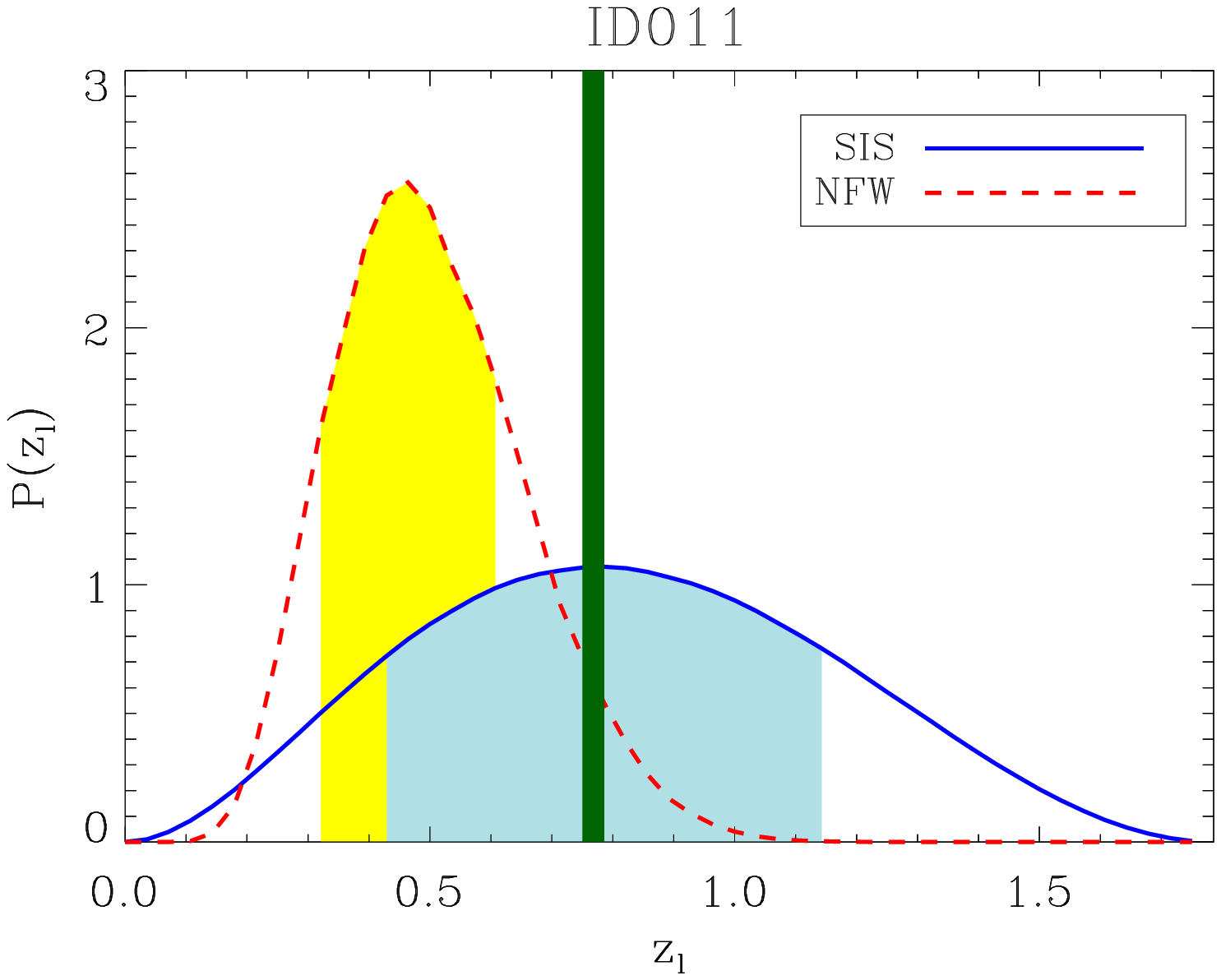}
  \includegraphics[width=0.49\linewidth]{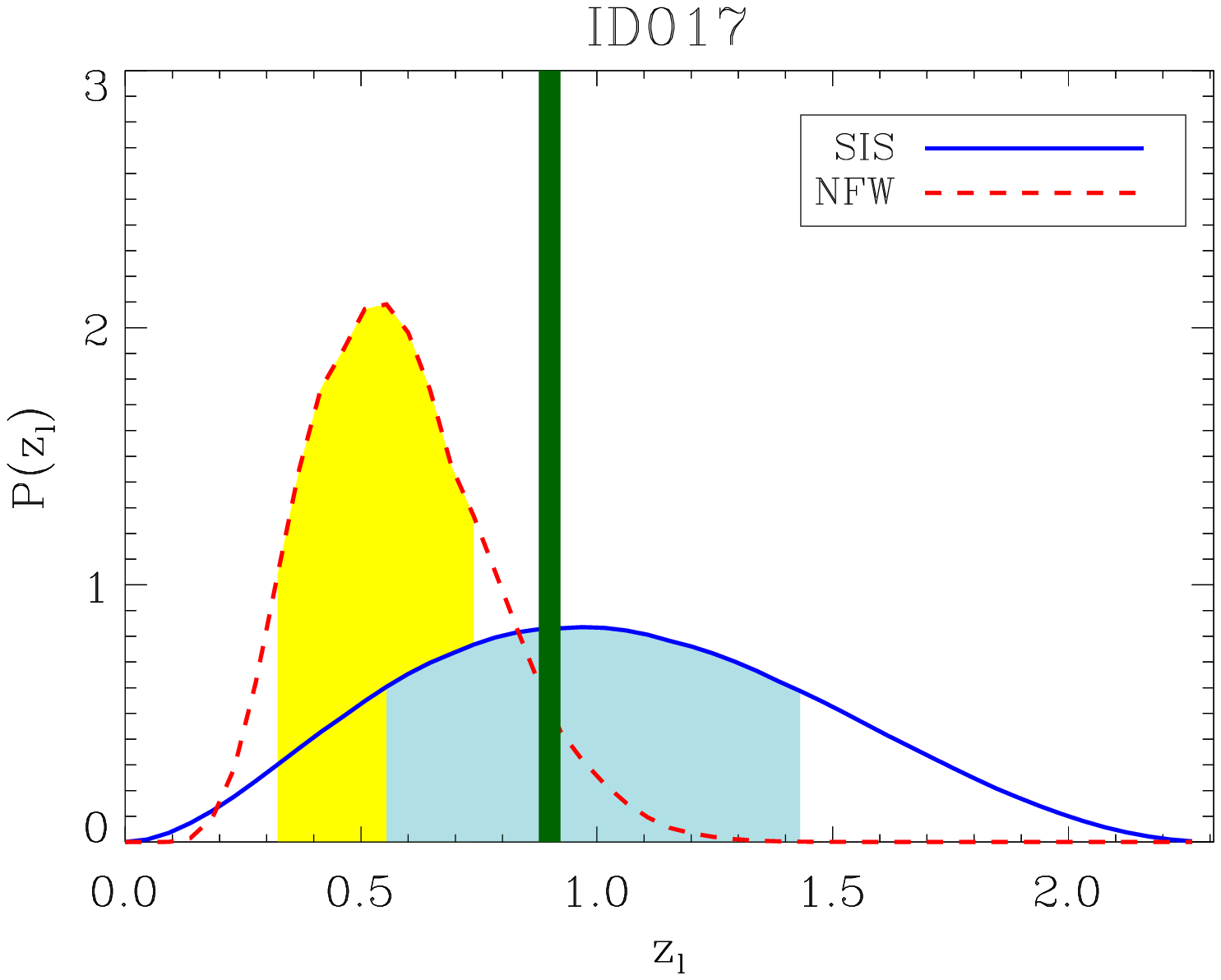}
  \includegraphics[width=0.49\linewidth]{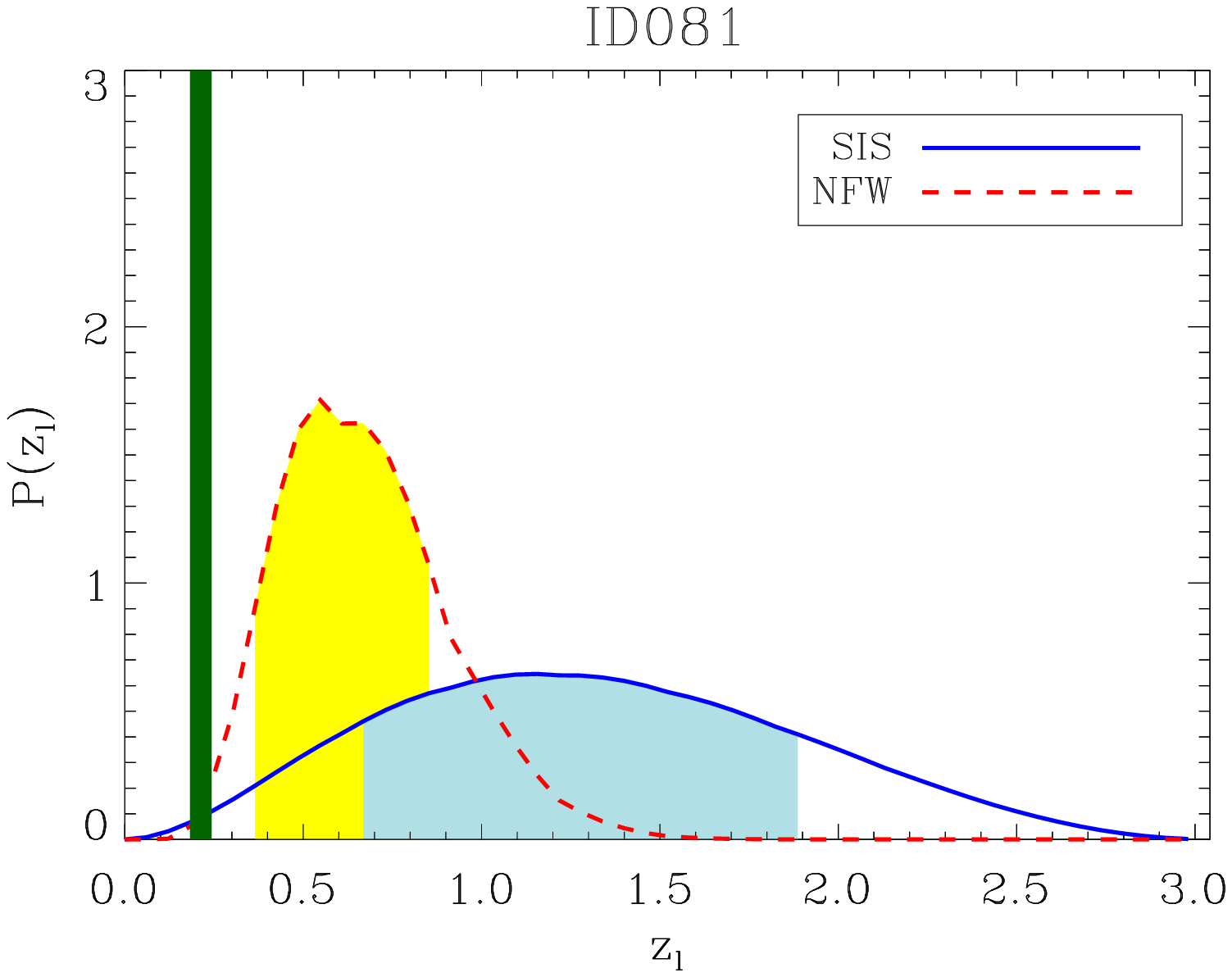}
  \includegraphics[width=0.49\linewidth]{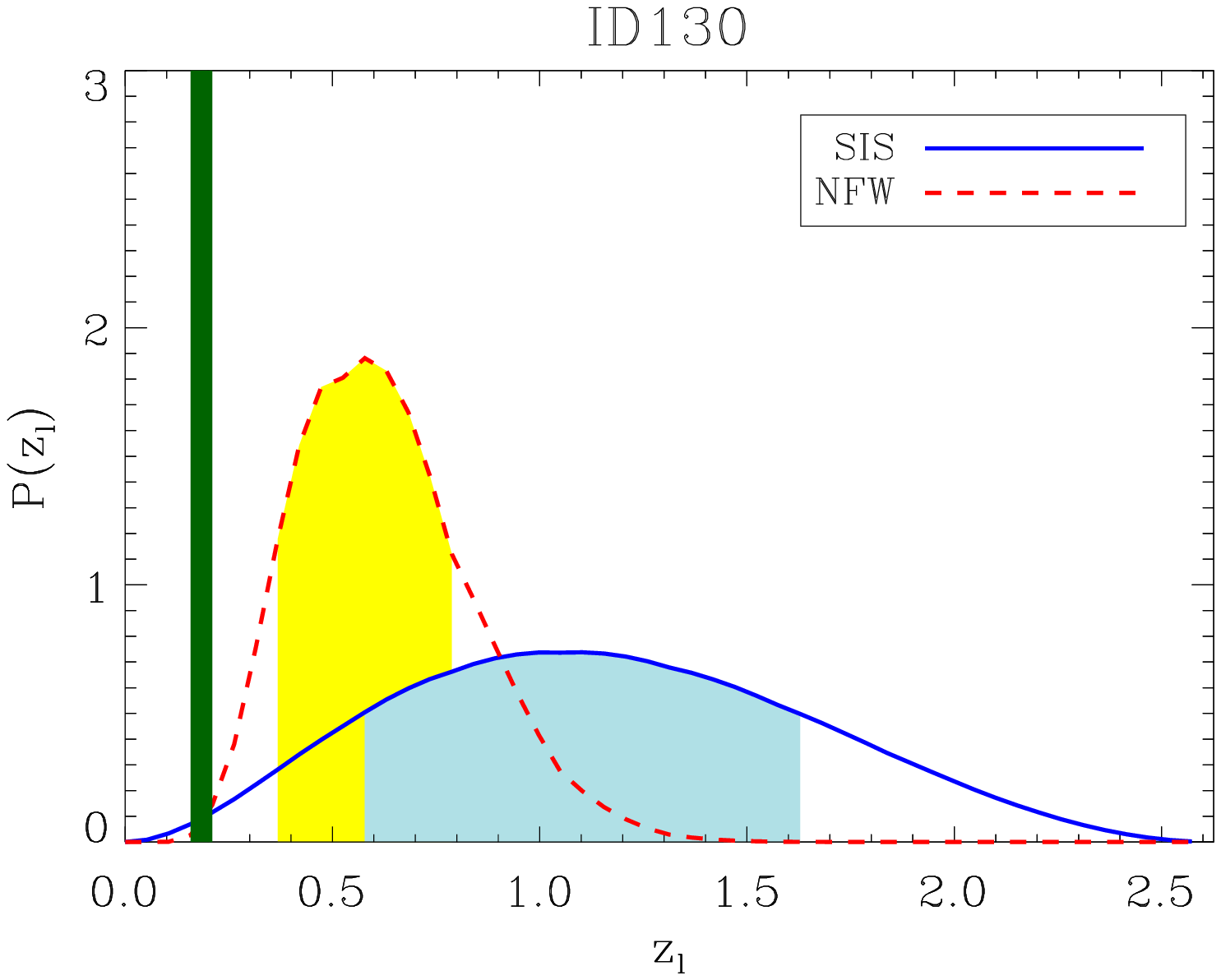}
  \caption{Conditional probability of finding a lens at a certain lens redshift given lensing has occurred, for all five \emph{Herschel}-ATLAS lenses. Source and lens redshifts are in Table \ref{tab1}. Solid blue lines use the SIS lens density profile; dashed red lines use the NFW density profile. Shading indicates 68\% confidence intervals. Solid green vertical lines indicate the estimated lens redshifts of the gravitational lenses found by \citet{Negrello2010}. }
 \label{figTauZl}
 \end{center}
\end{figure*}

Figure \ref{figTauZl} shows $\pzl$ for all five of the \emph{Herschel}-ATLAS SDP lenses. Again $\pzl$ was calculated as described in the Appendix assuming a \citeauthor{Tinker2008} mass function and a standard \lcdm cosmology, for any angular separation and magnification (i.e. $\theta>0$ and $\mu>1$). The source redshifts are as per the estimates in Table \ref{tab1}.

Three of the five \emph{Herschel}-ATLAS lensed sources (ID009, ID011, and ID017) show lens redshifts most consistent with 68\% confidence limits for the SIS density profile model.  These positions are still consistent with the NFW model at around the 95\% confidence level. The other two \emph{Herschel}-ATLAS lensed sources (ID081 and ID130) are both at much smaller lens redshifts. The $\zl$ are consistent with both the SIS and NFW models again at a 95\% confidence level. 

To quantify these results we can again calculate likelihood as:
\begin{eqnarray}
 \mathcal{L} \left( \zi \, \Big| \, \rho_\textsubscript{SIS/NFW}\right) = \prod\limits^{\nl}_{i=1} \,\pzi .
\label{eqn3}
\end{eqnarray}
Calculating this gives $\mathcal{L}\left(\zi \, | \, \rho_\textsubscript{SIS}\right) = 0.014$ and $\mathcal{L}\left(\zi \, | \, \rho_\textsubscript{NFW}\right) = 0.026$. So, in contrast to what was shown with $\pmu$, the NFW profile is preferred over the SIS density profile. 

Again we should note that there is likely to be an observational bias here. The way in which these lens systems are selected requires a lens object to be identified observationally. Objects which are closer will tend to be brighter and therefore easier to identify. Therefore it is likely that there is a selection bias for nearer lenses which has not been taken into account. This bias favours the NFW profile, so if it were taken into account the SIS profile would likely have a more favourable likelihood.

\subsection{Constraining the Dark Energy Density with $\dtzl$}
\label{sec2c}
Here we consider whether the dark energy density can be competitively constrained using the lens redshifts $\zl$ of the \emph{Herschel}-ATLAS SDP lenses as compared to the theoretical $\dtz2$. The tests above provide conflicting evidence for the SIS and NFW lens density profiles. Later in Section \ref{sec3} we will put these results into context using simulations, and project how many lensed sources are needed for a definitive result. Meanwhile, for illustrative purposes, here we assume a SIS density profile. Similarly since there is no information on the angular separation, we make the assumption that the mass function is well fit by a \citeauthor{Tinker2008} mass function.

Now we consider absolute probabilities we also need to take into account the other sources that were not lensed. This is to account for the probability of lensing occurring as well as the distribution of the lenses as a function of $\zl$. The likelihood is given by:
\begin{eqnarray}
 \mathcal{L} \left( \zl \,\Big|\,\oml \right) = \prod\limits^{\nl}_{i=1} \frac{d\tau\left(z_{\textsubscript{s}i},\zi\right)}{d\zl} \, \prod\limits^{N_\textsubscript{u}}_{j=1} \left(1-\tau\left(z_{\textsubscript{s}j}\right)\right),
\label{eqn4}
\end{eqnarray}
where $N_\textsubscript{u}$ is the number of unlensed sources. Given an estimated redshift distribution $\mathcal{N}(\zs)$ then, because the optical depth is small, we can approximate $\mathcal{L}\left(z_{\textsubscript{l}i}\,\Big|\,\oml \right)$ as \citep{Mitchell2005}:
\begin{align*}
 \ln \mathcal{L} \left(\zi \,\Big| \,\oml \right) =\,& \sum\limits^{\nl}_{i=1} \ln \frac{d\tau\left(z_{\textsubscript{s}i},\zi\right)}{d\zl} \, + \, \sum\limits^{N_\textsubscript{u}}_{j=1} \ln \left(1-\tau\left(z_{\textsubscript{s}j}\right)\right)
\\=\,& \sum\limits^{\nl}_{i=1} \ln \frac{d\tau\left(z_{\textsubscript{s}i},\zi\right)}{d\zl} \,- \, \sum\limits^{N_\textsubscript{u}}_{j=1} \tau\left(z_{\textsubscript{s}j}\right)
\\=\,& \sum\limits^{\nl}_{i=1} \ln \frac{d\tau\left(z_{\textsubscript{s}i},\zi\right)}{d\zl} - \int \mathcal{N}(\zs) \tau\left(\zs\right) d\zs.
\label{likelihood3}
\end{align*}
$z_{\textsubscript{s}i}$ and $\zi$ are the source and lens redshifts of the individual \emph{Herschel}-ATLAS  SDP lenses but $\zs$ in the second term is a discrete range from the minimum to maximum redshifts of \emph{Herschel}-ATLAS sources observed.
\begin{figure}
 \begin{center}
  \includegraphics[width=\linewidth]{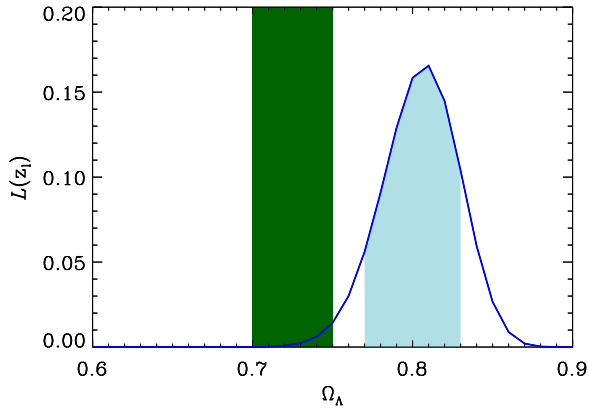}
  \caption{Likelihood of the observed lens redshifts of the \emph{Herschel}-ATLAS SDP lenses given a range of dark energy densities $\oml$. The shaded regions represent the 68\% confidence region. The vertical green line represents the current most favoured $\oml$.}
 \label{figTauzlcosmo}
 \end{center}
\end{figure}

The redshift distribution of the \emph{Herschel}-ATLAS sources is quite uncertain; for illustrative purposes here we assume that all the sources are at a redshift of 2. We take the number of unlensed sources as 252 from \cite{Clements2010}; the resulting $\oml$ is invariant to small changes in $N_\textsubscript{u}$. Figure \ref{figTauzlcosmo} shows the resulting likelihood as a function of the dark energy density (assuming a flat universe). This corresponds to an estimate of $\oml = 0.81^{+0.02}_{-0.04}$. This is consistent with the current best estimate for $\oml$ at the 95\% level.
\\\\
To summarise, there are some conflicting results, which is not surprising given the current limited number of confirmed gravitational lensed sources. In the next section we put these results in context using simulations. We also consider what constraints can be achieved with a larger number of lenses.

\section{Lens Statistics: Full \emph{Herschel}-ATLAS survey}
\label{sec3}
So far just five strong gravitational lenses have been confirmed in the \emph{Herschel}-ATLAS SDP data. With the full data set it should be possible to extract enough lensed sources with which to do more meaningful statistical analyses \citep{Negrello2010,Gonzalez-Nuevo2012}. Here we consider whether the full set of 100-1000 lenses will be sufficient to constrain the different models for the density profile, mass function and $\oml$ discussed above. 

The methodology is described in more detail below, but here is a general overview. The differential optical depth was calculated as described in the Appendix using different density profiles, mass functions and $\oml$. A thousand sets of $\nl$ lenses were generated by randomly sampling the differential optical depth at different redshifts $\zl$, magnifications $\mu$, and angular separations $\theta$. The corresponding likelihoods were then calculated as described in Equations \ref{eqn2}, \ref{eqn3} and \ref{eqn4} for each of the sets. The set size $\nl$ is varied to see how successfully the different approximations for $\rho$, $n$, and $\oml$ can be distinguished. Unless otherwise stated the source redshifts are all simply assumed to be $\zs = 2$. 

This work could be expanded to incorporate a full Bayesian analysis. It is not considered here as we do not consider a sophisticated enough halo model, and there is not yet enough data, to justify such an approach.   

\begin{figure}
 \begin{center}
  \includegraphics[width=\linewidth]{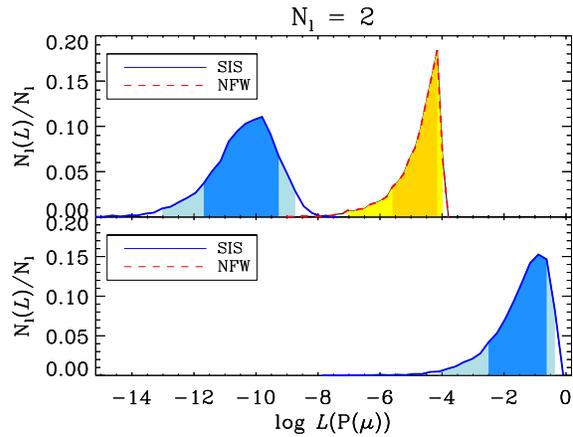}
  \caption{Normalised distribution of number counts of likelihoods realised from random sampling of theoretical probability distributions $\pmu$ for SIS (solid blue line) and NFW (dashed red line) density profiles. The plot is for sets of $\nl = 2$ lensed sources; the top plot is for lens redshifts generated assuming a NFW profile, and the bottom plot for SIS. The shading represents 68\% and 95\% confidence intervals. The two distributions only slightly overlap so it is fairly easy to distinguish between SIS and NFW using this approach - it will be even easier with more results.}
 \label{figTauzldp3}
 \end{center}
\end{figure}

First we look at constraining the lens density profile using $\pmu$. Ignoring the other possibilities (as discussed in Section \ref{sec1b}) here we focus on the SIS and NFW profiles. A thousand sets of $\nl = 2$ lens magnifications were generated by randomly sampling $\pmu$ for both SIS and NFW density profiles. The \citeauthor{Tinker2008} mass function and standard \lcdm parameters were assumed. Two versions of the 1000 sets were generated for each of the SIS and NFW density profiles. The associated likelihoods were calculated for each of these sets for both the SIS and NFW lens profiles as described in Eq. \ref{eqn2}. So in all four different combinations were considered. For each combination, the number counts for a range of different likelihoods were calculated. These results are shown in Figure \ref{figTauzldp3}. 

The blue and red curves are for likelihoods calculated with SIS and NFW density profiles respectively. The top plot is for sets of lens sampled from $\pmu$ using a NFW density profile; the bottom plot uses $\pmu$ for SIS density profiles. The two distributions only slightly overlap in the first instance, and not at all in the second case (the NFW distribution is at smaller $\mathcal{L}$ than on the plot). It is easy to distinguish between SIS and NFW density profiles because their likelihood are distinct even with this small amount of data. 

The actual likelihoods found from the SDP data above in Section \ref{sec2} were $\log\mathcal{L}\left(\mu_i \, | \, \rho_\textsubscript{SIS}\right) = -2.5$ and $\log\mathcal{L}\left(\mu_i \, | \, \rho_\textsubscript{NFW}\right)< -5.3$. These are both consistent with the bottom plot which gives likelihoods for lens redshifts sampled from $\pmu$ assuming a SIS density profile.

\begin{figure}
 \begin{center}
  \includegraphics[width=\linewidth]{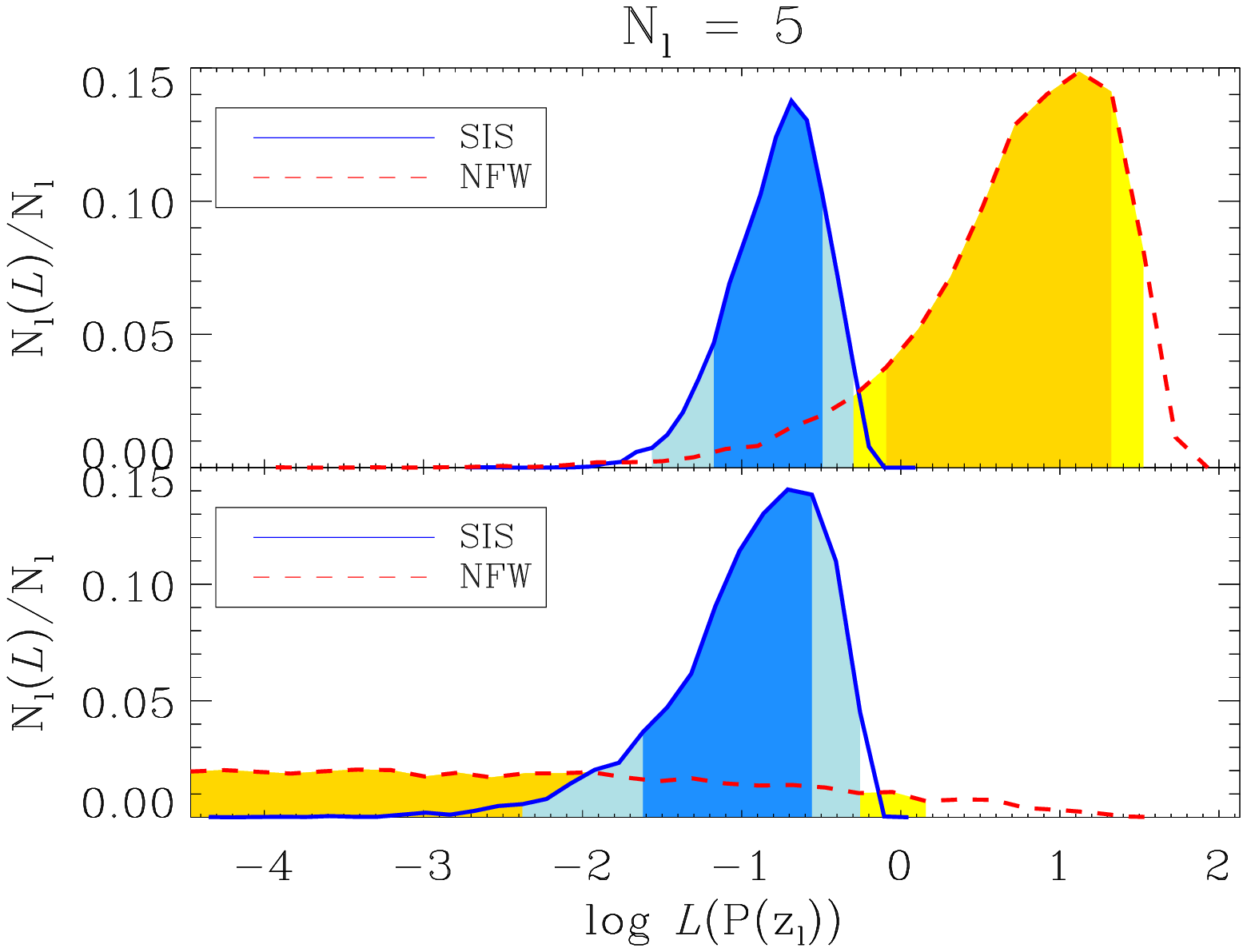}
  \includegraphics[width=\linewidth]{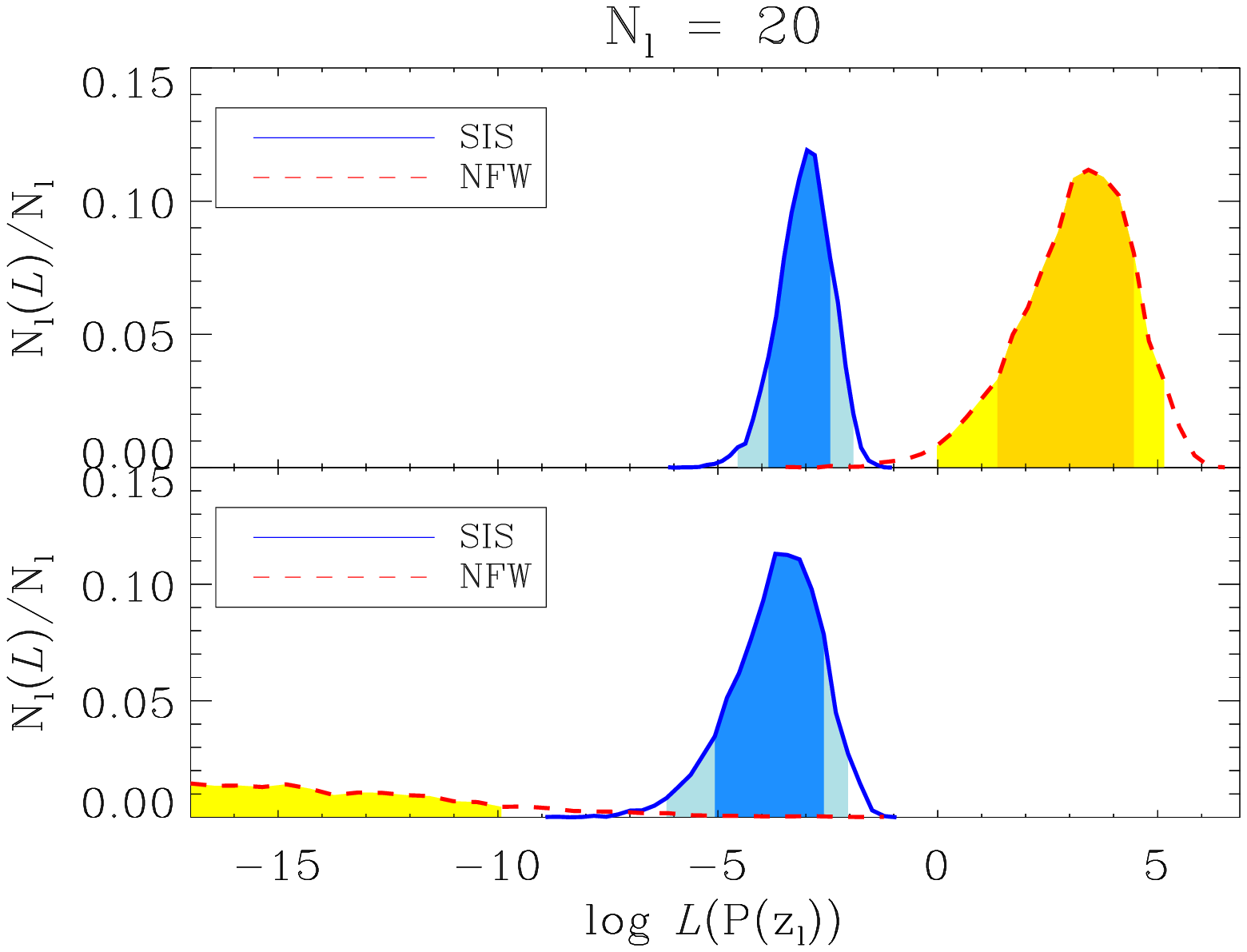}
  \caption{Normalised likelihoods realised from random sampling of theoretical probability distributions $\pzl$ given a \citeauthor{Tinker2008} mass function and standard \lcdm cosmological parameters, for SIS (blue solid line) and NFW (red dashed line) density profiles.  The top plot is for sets of $\nl = 5$ lensed sources (as we have for the current SDP sample); the lower plot is for $\nl = 20$ lenses. The top half shows the distribution of likelihoods found using SIS and NFW probabilities for lenses randomly sampled from the NFW $\pmu$. Similarly for the bottom plot but here the lenses are randomly sampled from the SIS $\pmu$. The shading represents 68\% and 95\% confidence intervals. The two distributions overlap for $\nl = 5$ so it is less easy to distinguish between SIS and NFW profiles; but at $\nl=20$ they are fairly easy to distinguish.}
 \label{figTauzldp2}
 \end{center}
\end{figure}

Whilst in Section \ref{sec2} $\pmi$ was found to be more consistent with a SIS model than a NFW model, $\pzi$ suggested the opposite.  Here the same test as above was performed again but this time sampling the lens redshift as opposed to the magnification. Figure \ref{figTauzldp2} shows the results. The top plot is for sets of $\nl = 5$ lensed sources (as we have for the current SDP sample). The top half shows the distribution of likelihoods found using SIS and NFW probabilities for lenses randomly sampled from the NFW $\pmu$. Similarly for the bottom plot but here the lenses are randomly sampled from the SIS $\pmu$. The two distributions overlap so it is less easy to distinguish between SIS and NFW using $\pzl$ than when using $\pmu$. Likelihoods found above in Section \ref{sec2} gave $\log\mathcal{L}\left(\zi \, | \,\rho_\textsubscript{SIS}\right)= -1.9$ and $\log\mathcal{L}\left(\zi \,| \,\rho_\textsubscript{NFW}\right) = -1.6$. Although the NFW likelihood is higher we see from Figure \ref{figTauzldp2} that is not inconsistent with a SIS lens. More data is needed to properly distinguish between them; $\nl=20$ lenses would be sufficient as seen from the bottom plot.  

Therefore to summarise, whilst the results for the density profile appeared conflicted initially in Section \ref{sec2}, here we see that actually they are both most consistent with a SIS density profile.

\begin{figure}
 \begin{center}
  \includegraphics[width=\linewidth]{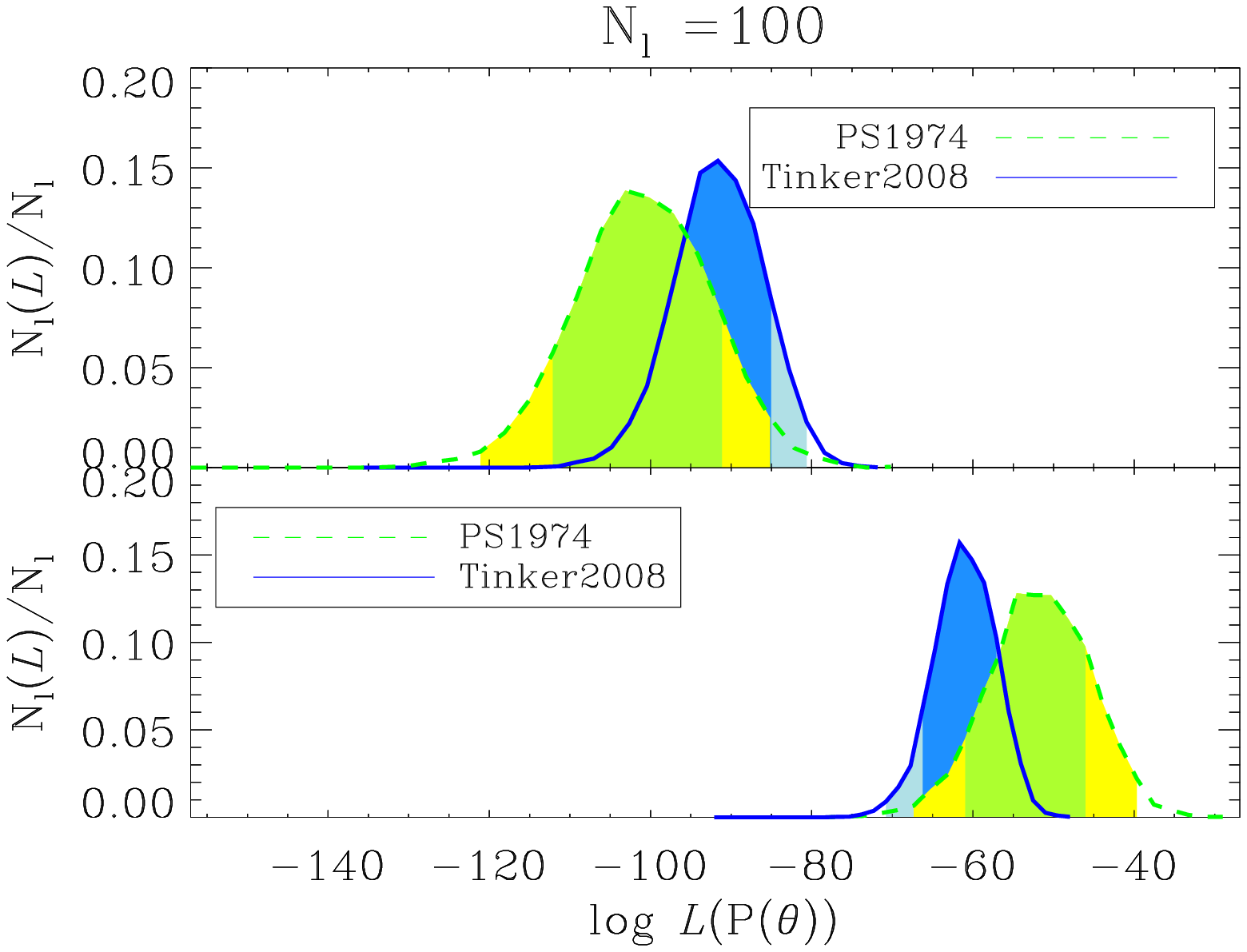}
  \caption{Normalised likelihoods realised from random sampling of theoretical probability distributions $\pth$ given a SIS density profile and standard \lcdm cosmological parameters, for Press-Schecter (green dashed line) and \citeauthor{Tinker2008} (blue solid line) mass functions. The plot is for sets of $\nl = 100$ lensed sources. The top half shows the distribution of likelihoods found using \citeauthor{Tinker2008} and \citeauthor{Press1974} mass function probabilities for lenses randomly sampled from the \citeauthor{Tinker2008} $\pth$. Similarly for the bottom plot but here the lenses are randomly sampled from the \citeauthor{Press1974} $\pth$. The shading represents 68\% and 95\% confidence intervals.  The distributions are not distinct so it is not possible to distinguish between the two mass functions using just the angular separations of 100 lensed images.}
 \label{figTauthetamf}
 \end{center}
\end{figure}

Values of the angular separation of lensed images are not yet available, but here we consider whether using $\theta$ would allow us to constrain the best fit mass function. Following the same approach as above but now sampling $\theta$ from $\pth$, the results for $\nl=100$ lenses are shown in Figure \ref{figTauthetamf}. This shows that even with angular separations for 100 lensed sources it would not be possible to distinguish between the \citeauthor{Press1974} and \citeauthor{Tinker2008} mass functions. Of the order 500 sources would be required as a minimum. The number of detections is expected to be this high, but the shear amount of follow-up that would be required to get enough angular separations $\theta$ makes this approach unfeasible.

Finally, we consider constraining the dark energy density $\oml$ with the absolute differential optical depth $\dtzl$. The lens redshifts are randomly sampled from the optical depth assuming the standard \lcdm cosmological parameters. Rather than considering two alternatives of $\oml$, here we consider a range of $\oml$ to calculate the likelihoods. Therefore rather than comparing the normalised distributions of $\mathcal{L}(\dtzl)$, instead we take the mean value $\bar{\mathcal{L}}$ for each $\oml$ and this is plot in Figure \ref{figTauzlcosmo}. The \citeauthor{Tinker2008} mass function and SIS density profile are used. The plot is for sets of $\nl = 5$ (as we have for the current SDP sample), $100$ and $1000$ lensed sources. The shaded areas are 68\% limits. 

We see that the SDP estimate of $\oml$ is consistent with the sort of errors expected for a sample of 5 lens redshifts, but with $\sim100$ lens redshifts the error on $\oml$ should be of order 0.01 i.e. comparable to the current best methods.

\begin{figure}
 \begin{center}
  \includegraphics[width=\linewidth]{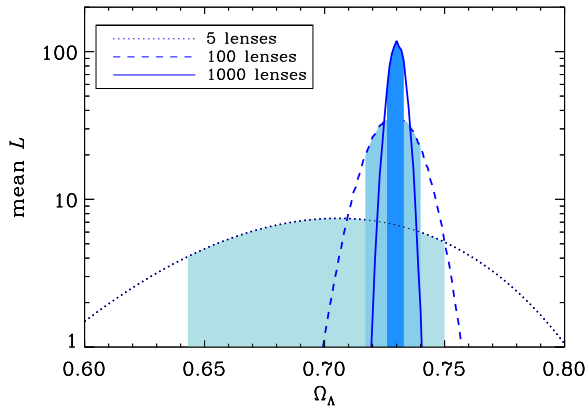}
  \caption{Mean likelihood as a function of dark energy density realised from random sampling of theoretical probability distributions $\dtzl$ given a SIS density profile and \citeauthor{Tinker2008} mass function. The shading represents 68\% confidence intervals. Assuming the other astrophysical uncertainties can be constrained, we see that competitive constraints on $\oml$ can be achieved with the full \emph{Herschel}-ATLAS data set.}
 \label{figTauzlcosmo2}
 \end{center}
\end{figure}

\section{Discussion and Conclusions}
Until recently the number of strong gravitational lenses that had been identified was relatively small for any particular survey. Now this situation is about to change dramatically as future surveys are expected to yield numbers of strong lenses into the hundreds and thousands. One of these surveys, \emph{Herschel}-ATLAS, expects to yield up to 1000 lenses. In this article we considered the statistics that can be done with this new data set of strong gravitational lenses.

First we reviewed the current analytical theory. We noted that the standard method for calculating magnification bias was not appropriate here. This is because the fast fall in number counts with flux in the submillimetre wavebands means that the standard calculation gives an infinitely large magnification bias. We therefore proposed a more realistic method. This effectively means the magnification bias is equal to the mean expected magnification of the gravitational lenses.

We saw that lensing statistics are sensitive to astrophysical properties, as well as the cosmological parameters, most significantly the lens density profile. Although so far only five lenses have been confirmed in the \emph{Herschel}-ATLAS SDP data, limiting its the statistical value, we have shown that these can be used to constrain the lens density profile. Initially the likelihoods calculated from each of $\pmu$ and $\pzl$ suggested conflicting results between the SIS and NFW lens density profiles. But simulations of the expected uncertainty of these likelihoods due to the small number of lenses showed that the results are perfectly consistent with a SIS, not a NFW, density profile. Of course neither of the SIS or NFW density profiles is probably realistic to explain the complex structure of the dark matter halo and baryon component. Here they are used to illustrate the effect of the current uncertainty in the density profile on the resulting optical depth.

We considered just one of the cosmological parameters for illustrative purposes. Whilst five lenses is not yet enough information to constrain the dark energy density, the full \emph{Herschel}-ATLAS data set should be more than sufficient to provide constraints competitive with other methods. How the mass function could be independently constrained was more uncertain and further work is clearly required to understand how this and other other uncertainties can most effectively be constrained. 

Here we have seen that the dominant uncertainties in gravitational lens statistics are not cosmological but astrophysical i.e. understanding the nature of the lensing objects. Whilst more investigation is required, we can see that new data sets of strong gravitational lenses will be fundamental for constraining both halo astrophysics as well as cosmological parameters.

\section*{Acknowledgments}
Many Thanks to Loretta Dunne and Steve Maddox for useful discussions. For the purposes of this work Jo Short is supported by STFC STEP fellowship, Elizabeth Pearson received funding from a STFC studentship, Peter Coles is partly supported by STFC Standard Grant ST/G002231/1, and Steve Eales and Peter Coles are supported by STFC Rolling Grant ST/H001530/1.

\appendix
\section{Theoretical Statistics of Gravitational Lenses }
\label{app1}
\subsection{General Formulation}
A source is defined to be \emph{strongly gravitationally lensed} when it has been multiply imaged due to the gravitational effects of mass between the source and the observer. The probability of a source being lensed is described by the optical depth ($\tau$). It depends on three main components: the \emph{cosmology} which determines the co-moving volume element at a given redshift; the \emph{normalised mass function} ($n$) which describes the number density of halos; and the \emph{lensing cross-section} ($\sigma$), which is the area in the lens plane where the separation between the lens and source is sufficiently small for strong lensing to occur:
\begin{eqnarray}
 \tau(\zs) =   \int\limits^{\zs}_0 \int\limits^\infty_0 \, \frac{d\dl}{d\zl}\, (1+\zl)^3 n(m,\zl)\, \sigma(m,\zl)\, dm\, d\zl,
\label{eqnTau1}
\end{eqnarray}
where $m$ is the mass of the lens, $\zl$ is the redshift of the lens, $\zs$ is the redshift of the source, and $\dl$ is the angular diameter distance from the observer to the lens \citep{Schneider1993}. 
Below we describe how each of these components can be calculated.

\subsection{Angular Diameter Distance}
The angular diameter distance\footnote{This definition assumes a flat universe.} between the two redshifts $z_a$ to $z_b$ is given by:
\begin{eqnarray}
D_{ab} = \frac{c}{H_0}\frac{1}{1+z_b}\int\limits_{z_a}^{z_b}  E(z) dz,
\end{eqnarray}
where $ E(z)  =[ \omm(z=0)(1+z)^3 + \oml(z=0)]^{0.5}$. 

\subsection{Mass Function}
\label{secApMassFunc}
The differential number density of halos ($n$, sometimes written as $dn/\,dm$) is commonly estimated using this expression derived by \cite{Press1974}:
\begin{eqnarray}
  \frac{m^2 n(m,\zl)}{\bar\rho}\frac{dm}{m} =\nu f(\nu)\frac{d\nu}{\nu}.
\end{eqnarray}
$\bar\rho=\omm\rho_\textsubscript{c}$ is the average matter density, $ \nu(m,\zl) = \delta_\textsubscript{c}^2/\sigma^2(m,\zl)$, $\delta_\textsubscript{c}= 1.686$ is the critical density required for spherical collapse and $\sigma^2$ is the variance in the linear density fluctuation field\footnote{We use the \cite{Bardeen1986} transfer function to calculate the linear power spectrum.}. Two approximations for $f(\nu)$ are those by \cite{Press1974} and \cite{Tinker2008}:
\begin{eqnarray}
   f_\textsubscript{PS}(\nu) = \sqrt{\frac{1}{2\pi\nu}}\ \exp\left(\frac{-\nu}{2}\right),
  \label{fps}
\end{eqnarray}
and
\begin{eqnarray}
  f_\textsubscript{T}(\nu) = A \left[ \left( \frac{\delta_\textsubscript{c}}{b\sqrt{\nu}} \right) ^{-a} +1 \right] \exp\left(\frac{-\nu c}{\delta_\textsubscript{c}^2}\right),
  \label{ftinker}
\end{eqnarray}
where $A = 0.186(1+z)^{-0.14}$, $a = 1.47(1+z)^{-0.06}$, $b = 2.57(1+z)^{-\alpha}$, $\alpha = 1.068$, and $c = 1.19$.

\subsection{Lensing Cross-section}
The gravitational lensing cross-section ($\sigma$) is the area in the lens plane where the lens and source are within the critical angular separation $\beta_\textsubscript{c}$ for strong lensing to occur \citep{Schneider1993}:
\begin{eqnarray}  \sigma = \pi \beta_\textsubscript{c}^2 \dl^2. \label{eqnCrossSec}\end{eqnarray} 
The critical angular separation $\beta_\textsubscript{c}$ that produces $\sigma$ depends on the shape of the lens density profile. Two different approximations for the lens density profile (SIS and NFW) are considered in Sections \ref{secSIS} and \ref{secNFW}. First the general approach for spherically symmetric lens density profiles is reviewed.

\begin{figure}
 \begin{center}
  \includegraphics[width=0.95\linewidth]{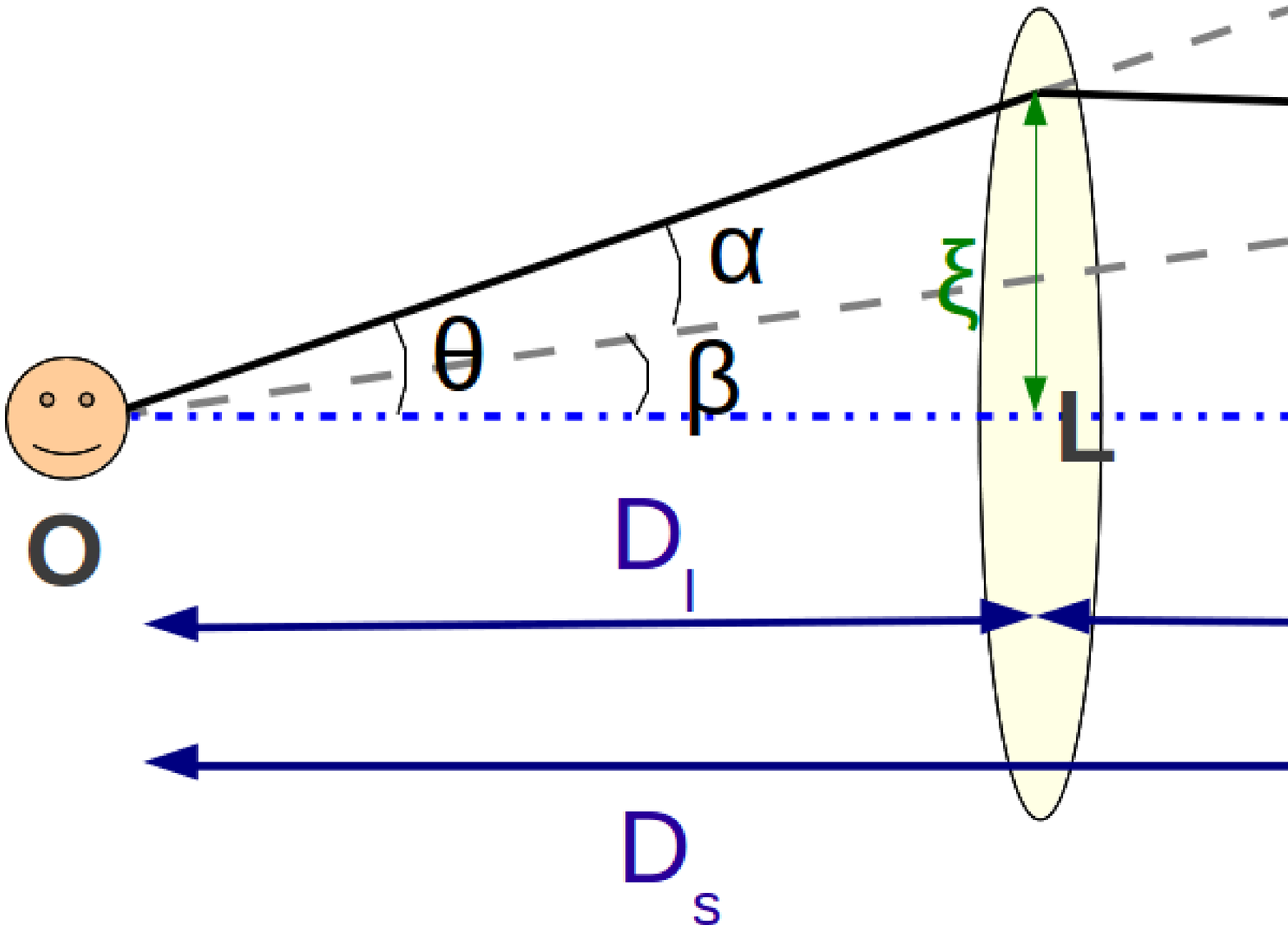}
  \caption{Illustrative diagram of a gravitational lens system. The source, lens and observer are positioned at points S, L and O respectively. The source and lens are at distances $D_\textsubscript{s}$ and $\dl$ from the observer, and separated by a distance $D_\textsubscript{ls}$. $\xi$ and $\eta$ are distances in the lens and sources plans from the observer-lens axis (dot-dash blue line).  The angular separation of the source from the observer-lens axis, as would be seen by the observer without lensing, is $\beta$. However the observer does not see the source directly but a modified image (I) of the source due to the deflection by the lens. This image appears to be at an angular separation from the observer-lens axis of $\theta$. The actual deflection angle $\hat{\alpha}$, and the reduced deflection angle $\alpha$, can be estimated using the thin lens approximation as detailed in Eq. \ref{eqnAlpha}.}
  \label{FigGLensSys}
 \end{center}
\end{figure}

The critical angle ($\beta_\textsubscript{c}$ in Eq. \ref{eqnCrossSec}) can be estimated using the \emph{lens equation}:
\begin{eqnarray}
\beta = \theta - \alpha,
\end{eqnarray}
where $\alpha = \hat{\alpha} D_\textsubscript{ls} / D_\textsubscript{s}$  (see Figure \ref{FigGLensSys}). Note this assumes a flat space time so that the Euclidean relation \emph{separation = angle $\times$ distance} holds.

The lens equation can alternatively be written in terms of the radial distances, $\eta / D_\textsubscript{s} = \xi / \dl - \hat\alpha D_\textsubscript{ls} / D_\textsubscript{s}$. This can be simplified further using dimensionless distances $ y = \eta / \eta_0 $ and $x=\xi / \xi_0$ where $\eta_0=\xi_0 D_\textsubscript{s} / \dl$, to give the lens equation of the form:
\begin{eqnarray} y = x - \hat\alpha\frac{D_\textsubscript{ls}\dl}{D_\textsubscript{s}\xi_0}. \label{eqnLens} \end{eqnarray}

Using the \emph{thin lens approximation},  which uses the fact that the lens is thin compared to the total path traveled by the light, the deflection angle $\hat\alpha$ can be calculated as:
\begin{eqnarray}  \hat{\alpha}(x) = \frac{4Gm(x)}{x \xi_0 c^2}, \label{eqnAlpha} \end{eqnarray}
where $m(x)= 4\pi\int\nu\int\rho(r) dz\, d\nu$ and $r^2/\rs^2 = z^2 + x^2$.

The distance in the lens plane subtended by the critical angle is simply $y_\textsubscript{c} = \beta_\textsubscript{c} D_\textsubscript{s}/\eta_0$. This is called the critical radius and is calculated as the value of $y(x)$ where $dy/dx=0$. For values of $y$ below $y_\textsubscript{c}$, multiple images are formed. We will see below that two multiple images are formed in the case of the SIS density profile, and three in the case of the NFW density profile. 
 
\subsection{Singular Isothermal Sphere (SIS) density profile}
\label{secSIS}
The SIS density profile is given by:
\begin{eqnarray}   \rho(r) = \frac{v^2}{2\pi G r^2}. \end{eqnarray}
The velocity dispersion\footnote{Note, this velocity dispersion is from SDSS estimates for the CLASS survey.} can be approximated as:
\begin{eqnarray}
 v(m,\zl) = 92.3 \left[\Delta^{0.5}_{vir}(\zl) E(\zl) \frac{m}{10^{13} h^{-1} M_\odot}\right]^{1/3} \textnormal{km/s},
\end{eqnarray}
where $ \Delta_{vir}(\zl) = 18\pi^2 + 60[\Omega(\zl) - 1] - 32[\Omega(\zl) -1]^2 $ and $\Omega(\zl) =\omm(0)(1+\zl)^3/E(\zl)^2$ \citep{Mitchell2005}.

The deflection angle (Eq. \ref{eqnAlpha}) becomes $\hat\alpha =4\pi v^2/c^2$ and the lens equation (Eq. \ref{eqnLens}) reduces to  $ y = x \pm 1 $ where:
\begin{eqnarray}
     \xi_0 = \frac{4\pi v^2}{c^2}  \frac{D_\textsubscript{ls}\dl}{D_\textsubscript{s}}.
\end{eqnarray}

\begin{figure}
 \begin{center}
  \includegraphics[width=\linewidth]{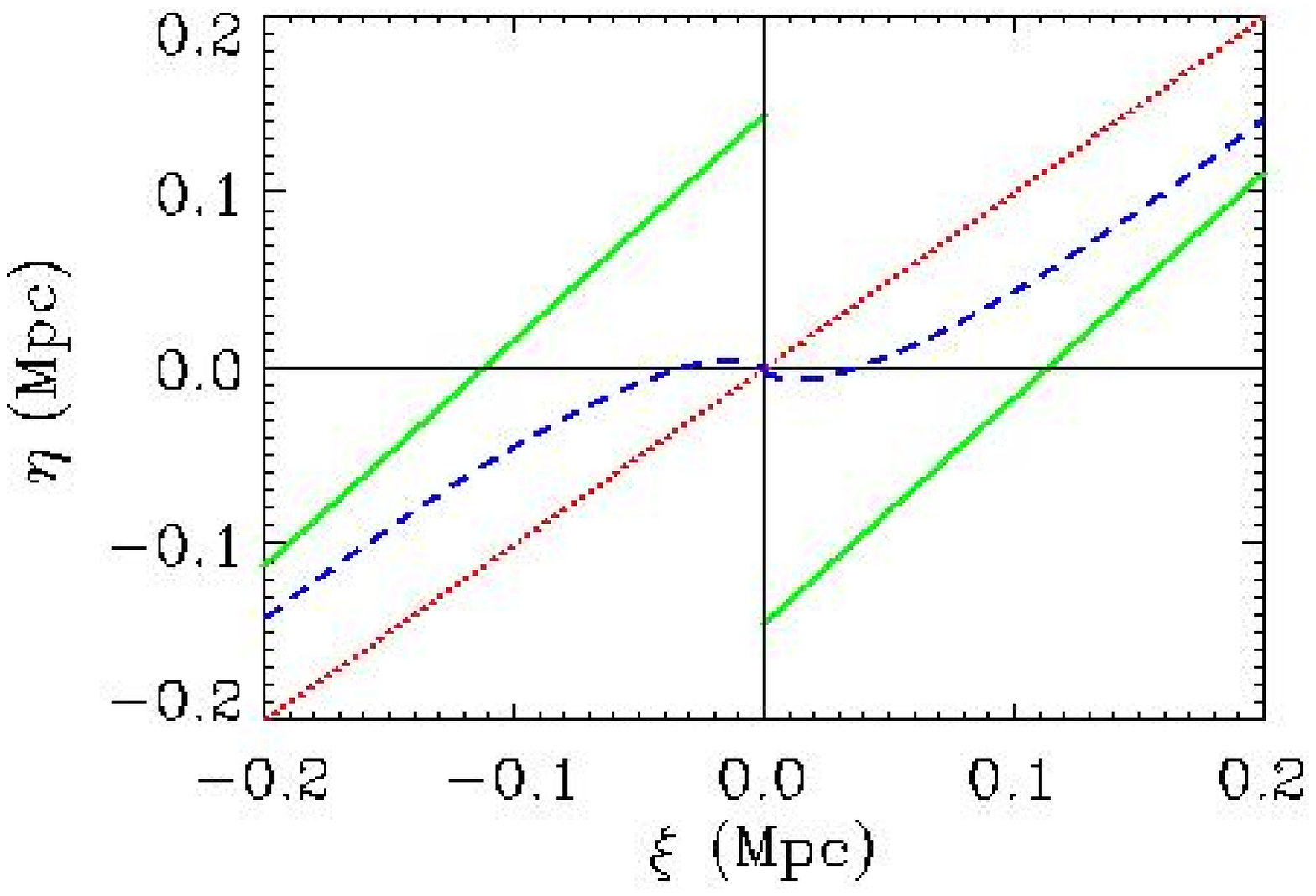}
   \caption{Source position ($\eta$) as a function of lens position ($\xi$) from the lens equation (Eq. \ref{eqnLens}) for SIS (green solid line) and NFW (blue dashed line) profiles. The diagonal red dotted line shows the case where there is no lensing. This example is for $\zs = 3$, $\zl = 0.5$ and $m = 10^{15}$ M$_\odot$. For the NFW lenses, strong lensing (generation of multiple images) occurs for source positions with magnitudes less than the local extrema. For SIS lenses, there are no local extrema but here strong lensing occurs for $\eta < |\eta(0)|$.  }
  \label{figLensEqn}
 \end{center}
\end{figure}

The critical value of $y$ below which strong lensing (or multiple images) occurs is $|y_c|= 1$. We see from Figure \ref{figLensEqn} that two images are generated in this case. The lensing cross-section is given by:
\begin{eqnarray}
 \sigma_\textsubscript{SIS} = \pi \beta_c^2 \dl^2 =  \frac{\pi y_c^2 \eta_0^2\dl^2}{D_\textsubscript{s}^2}  =\pi \left[\frac{4\pi v^2 D_\textsubscript{s} D_\textsubscript{ls}}{ c^2 \dl}\right]^2.
\end{eqnarray}

Other properties of the lenses, such as the magnification and angular separation, can also be calculated. The probability that a lens will be magnified more than $\mu$ is included in the cross-section as:
\begin{eqnarray}
  \sigma (>\mu) = \sigma A(>\mu),
\label{eqnCsMag}
\end{eqnarray} 
where $A(>\mu)$ describes the fractional area satisfying magnifications greater than $\mu$:
\begin{eqnarray}
  A(>\mu) = \frac{1}{\pi y_\textsubscript{c}^2}\int\limits^{y_\textsubscript{c}}_{-y_\textsubscript{c}} \pi y\, \Theta[\mu(y) - \mu]\, dy ,
  \label{eqnMagFactor}
\end{eqnarray} 
where $\Theta$ is a step function. The magnification of an image is given by:
\begin{equation}
  \mu(y)=\left|\frac{y}{x}\frac{dy}{dx}\right|^{-1},
  \label{eqnMagY}
\end{equation}
\citep{Schneider1993} and since the solutions of the SIS lens equation are  $x_\pm=y\pm 1$ ($|y|<1$), this becomes:
\begin{equation}
  \mu_\pm(y)=\frac{1}{y}\pm 1.
\end{equation}
So the total magnification $\mu(y) =  \mu_-(y) + \mu_+(y) = 2/y$. Substituting this into Eq. \ref{eqnMagFactor} gives:
\begin{eqnarray}
  A(>\mu) = \frac{1}{\pi y_\textsubscript{c}^2}\int\limits_{-y_\textsubscript{c}}^{y_\textsubscript{c}} \pi y\, \Theta[\frac{2}{y}- \mu]\, dy
	  =\frac{4}{\mu^2} \textnormal{ for } \mu >2.
\end{eqnarray} 
Therefore $ \sigma_\textsubscript{SIS} (>\mu) =  4 \sigma_\textsubscript{SIS}\mu^{-2}  $ for $\mu > 2 $. 

Similarly the probability that a lens will have images with an angular separation greater than  $\theta$ is included in the cross-section as:
\begin{eqnarray}
  \sigma_\textsubscript{SIS} (>\theta) = \sigma_\textsubscript{SIS} \Theta(\vartheta -\theta).
\end{eqnarray}
The separation between the two images can be seen from the lens equation to be $\Delta x = 2$. The angular separation becomes $\vartheta = \xi_0 \Delta x / \dl= 8 \pi v^2 D_\textsubscript{ls}/c^2 D_\textsubscript{s}$.
				  
\subsection{\citeauthor{Navarro1997} (NFW) density profile}
\label{secNFW}
\cite{Navarro1997} proposed a halo density profile of the form:
\begin{eqnarray}
\rho(r)=\frac{\rhos}{\left(\frac{r}{\rs}\right)\left(1 + \frac{r}{\rs}\right)^2},
\end{eqnarray}
where $\rs$ is the scale radius and $\rhos$ is the density at $\rs$. The scale radius $\rs$ can be calculated as $\rs = r_\textsubscript{vir} / c$ where  the virial radius ($r_\textsubscript{vir}$) is defined as the radius at which the mean density of the mass is 200 times that of the matter density. The concentration can be approximated using \citep{Bullock2001}:
\begin{eqnarray}
c(m,\zl) = \frac{9}{1 + \zl} \Big( \frac{m}{1.5\times10^{13}M_\odot} \Big)^{-0.13},
\end{eqnarray}
and $\rhos$ can be calculated using, $ m = 4 \pi \rhos \rs^3 [ \ln(1+c)  - c / (1 + c) ].$
The lens equation (Eq. \ref{eqnLens}) becomes:
\begin{eqnarray}
 y = x -\frac{\mu_\textsubscript{s}}{x} g(x),
\end{eqnarray}
where  $\xi_0 = \rs$, $\mu_\textsubscript{s} = 4\rhos \rs/\Sigma_\textsubscript{crit}$. The critical surface density is given by:
\begin{eqnarray}
 \Sigma_\textsubscript{crit} = c^2 D_\textsubscript{s}/4\pi G \dl D_\textsubscript{ls},
\end{eqnarray}
and:
\begin{eqnarray}
 g(x) =\int\limits_0^x u  \int\limits_0^\infty (u^2+v^2)^{-0.5}(1 + (u^2+v^2)^{0.5})^{-2} dv\, du.
\end{eqnarray}

It can be seen from Figure \ref{figLensEqn} that the critical value of $y$ below which strong lensing (or multiple images) occurs is when $dy/dx=0$. So the cross-section is given by:
\begin{eqnarray}
 \sigma_\textsubscript{NFW}(\zs, \zl, m) = \pi \beta_\textsubscript{c}\dl^2 = \pi \left( \frac{y_\textsubscript{c}\eta_0}{D_\textsubscript{s}} \right) \dl^2= \pi y_\textsubscript{c}^2 \rs^2.
\end{eqnarray}

As for the SIS profile, the dependence of $\sigma$ on other factors such as magnification and angular separation can be calculated. The lensing cross-section can be calculated in the same way as described in Equations \ref{eqnCsMag} and \ref{eqnMagFactor}. In the case of the NFW profile this does not reduce into a simple expression as is the case for the SIS. Instead $\sigma_\textsubscript{NFW}(>\mu)$ needs to be calculated using the full expression detailed in Equations \ref{eqnCsMag} to \ref{eqnMagY}. This calculation is extensive and a good approximation is available from \cite{Oguri2002}.

Again to calculate the dependence of the cross-section of the angular separation, $ \sigma_\textsubscript{NFW} (>\theta) = \sigma_\textsubscript{NFW} \Theta(\vartheta -\theta)$. The separation between the two furthest images can be seen from the lens equation to be $\Delta x = 2 x_\textsubscript{c}$ where $x_\textsubscript{c} = \beta_\textsubscript{c}D_\textsubscript{s}$. The angular separation  is therefore $\vartheta = \xi_0 \Delta x / \dl= 2 \rs x_\textsubscript{c} / \dl $.

\end{document}